\documentclass[12pt,preprint]{aastex}
\usepackage{lscape} 
\newcommand{\co}{\mbox{$^{12}$CO}}
\newcommand{\coa}{\mbox{$^{13}$CO~}}

\newcommand{\kms}{\mbox{km s$^{-1}$}}
\newcommand{\htwo}{\mbox{H$_2$}}

\newcommand{\cc}{\mbox{cm$^{-3}$}}

\newcommand{\msun}{\mbox{M$_\odot$}}
\newcommand{\mpcsq}{\mbox{M$_\odot pc^{-2}$}}
\newcommand{\vlsr}{\mbox{$V_{LSR}$}}

\shorttitle{Galactic Molecular Clouds}
\shortauthors{Heyer et al.}

\begin{document}
\title{Re-examining Larson's Scaling Relationships in Galactic Molecular Clouds}

\author{Mark Heyer\altaffilmark{1}, Coleman Krawczyk\altaffilmark{1,2}, 
 Julia Duval\altaffilmark{3}, James M. Jackson\altaffilmark{3}}

\altaffiltext{1}{Department of Astronomy, University of Massachusetts, Amherst,
MA 01003-9305; heyer@astro.umass.edu}
\altaffiltext{2}{Department of Physics, Drexel University, Philadelphia, PA 19104} 
\altaffiltext{3}{Institute for Astrophysical Research, Boston University, Boston, MA 02215}

\begin{abstract}
The properties of Galactic molecular clouds tabulated by Solomon et~al. 
(1987) (SRBY) are re-examined using the 
Boston University-FCRAO Galactic 
Ring Survey of \coa\ J=1-0 emission.  
These new data 
provide a lower opacity tracer of molecular clouds and 
improved angular and spectral resolution compared to previous surveys of 
molecular line emission along the Galactic Plane.  
We calculate GMC masses within the SRBY cloud boundaries
assuming LTE conditions throughout 
the cloud and a constant \htwo\ to \coa\ abundance, while 
accounting for the variation of the $^{12}$C/$^{13}$C 
with Galacto-centric radius.  The LTE derived masses are typically five times 
smaller than the SRBY virial masses.  The corresponding 
median mass surface density of molecular hydrogen for this sample 
is 42 \mpcsq, which is significantly lower than 
the value derived by SRBY (median 206 \mpcsq) 
that has been widely adopted by most models 
of cloud evolution and star formation.  This discrepancy arises from 
both the extrapolation by SRBY of velocity dispersion, size, and CO 
luminosity to the 1~K 
antenna temperature isophote that likely overestimates the GMC masses 
and our assumption of constant \coa\ abundance over the projected 
area of each cloud.  Owing to the uncertainty of molecular abundances in the 
envelopes of clouds, the mass surface density of giant molecular clouds 
could be larger than the valued derived from our \coa\ measurements.  
From velocity dispersions derived from the \coa\ data, 
we find that the coefficient of the cloud structure 
functions, $v_\circ=\sigma_v/R^{1/2}$, is not constant, as required to 
satisfy Larson's scaling relationships, but rather systematically varies with 
the surface density of the cloud as $\sim\Sigma^{0.5}$ as expected for 
clouds in self-gravitational equilibrium.  
\end{abstract}
\keywords{ISM: clouds -- ISM: kinematics and dynamics
}

\section{Introduction}
Giant molecular clouds (GMCs) are the exclusive sites of 
star formation in galaxies.  Their evolution 
and conversion 
of interstellar material into stars are governed by the 
interplay between self-gravity, magneto-turbulent pressure, and feedback 
processes from newborn stars (McKee 1989). The configuration of a GMC,
parameterized by its observed size, velocity dispersion, and 
mass surface density, 
offers a snapshot view of its dynamical state.
Larson (1981) identified scaling 
relationships between these observable quantities that have provided 
the basic, grounding point for all subsequent descriptions of 
interstellar molecular clouds and star formation These scaling 
relationships are: 1) a power relationships  between the velocity 
dispersion, $\sigma_v$, and the spatial scale of the emitting volume, $L$, $\sigma_v \sim L^{0.38}$, 
2) self-gravitational equilibrium, $2\sigma_v L^2/GM\sim 1$, and 3) an inverse relationship between 
the mean density, $n$, and size of the cloud, $n \sim L^{-1.1}$.  The last of these 
relationships implies that all molecular clouds have comparable gas surface density. 
The original compilation of these scaling relationships used molecular 
line data available from earlier studies.
Many of these data were collected with the 
earliest millimeter wave telescopes and instrumentation  and 
included spatially undersampled maps of molecular line emission from a limited number of nearby ($<$ 2.2 kpc) interstellar clouds with poor sensitivity, compared to currently available data.

The Larson scaling relationships have been supplemented 
with additional observations with improved sensitivity and larger 
samples.  The most significant study to confirm Larson's 
results is described by Solomon et~al. (1987) (hereafter, SRBY). 
They used the University of Massachusetts-Stony Brook (UMSB)
Galactic Plane Survey
(Sanders et~al. 1985) to identify 273 GMCs in the first quadrant. 
Each cloud was 
defined as a closed surface within the longitude-latitude-velocity 
data cube at a given threshold of antenna temperature. For most entries 
in the catalog, 
the threshold was 4 K (T$_R^*$).  This high threshold, relative 
to the noise of the data, was necessary to avoid the blending of 
emission at lower intensity levels from unrelated clouds that are 
densely distributed within the $l-b-V_{LSR}$ domain of the 
spectroscopic observations.
The blending is particularly severe near the 
tangent points at each Galactic longitude.  Realizing 
that such a high threshold would not fully account for the
bulk of the emission from a GMC, SRBY extrapolated the sizes, 
velocity dispersions, and CO luminosities to the 
1 K isophote.
After accounting for this low level contribution and its
effect on the tabulated properties, SRBY identified a 
size-linewidth relationship with a steeper power relationships index (0.5) than 
derived by Larson (1981) and concluded that GMCs are  
self-gravitating objects in virial equilibrium.
An algebraic consequence of these two results is that the 
molecular gas surface density is constant for all 
clouds with $\Sigma(H_2)=200$ \mpcsq.
This value is corrected from 170 \mpcsq\ quoted in SRBY 
to account for the difference in the values adopted for 
the Galactocentric radius of the Sun (10 kpc vs 8.5 kpc) that 
affects the virial mass. 
We have similarly corrected other values in the SRBY catalog
(CO luminosity, distance, galactocentric radius, sizes)
using the rotation curve of Clemens (1985) with $R_\odot$=8.5 kpc.

The data used by Larson (1981) and SRBY are not optimal for 
deriving properties of interstellar clouds.  In  both cases,
the observed cloud fields were spatially undersampled with respect to 
the angular resolution of the telescope used to gather the CO data. Owing to 
poor sensitivity or the need for identifying clouds at 
a high intensity threshold, the target clouds are 
biased towards the brightest interstellar clouds, the brightest 
regions within the clouds,  or clouds 
that happen to be nearby or associated with conspicuous 
star formation. In the case of SRBY, the tracer of molecular 
gas was \co\ J=1-0 line emission that is optically thick 
under most prevailing conditions in molecular clouds. 
Given these limitations, it is reasonable to inquire whether the 
properties of GMCs and the corresponding 
Larson scaling relationships can withstand scrutiny with 
vastly superior data available today.  For example, a recent 
study by Bolatto et~al. (2008) examined the properties
of resolved 
giant molecular clouds in dwarf galaxies and those within Local Group 
spiral galaxies.  They found the properties of extragalactic GMCs similar 
to those determined by SRBY for Galactic GMCs.   

\begin{figure}
\begin{center}
\includegraphics[width=0.7\hsize]{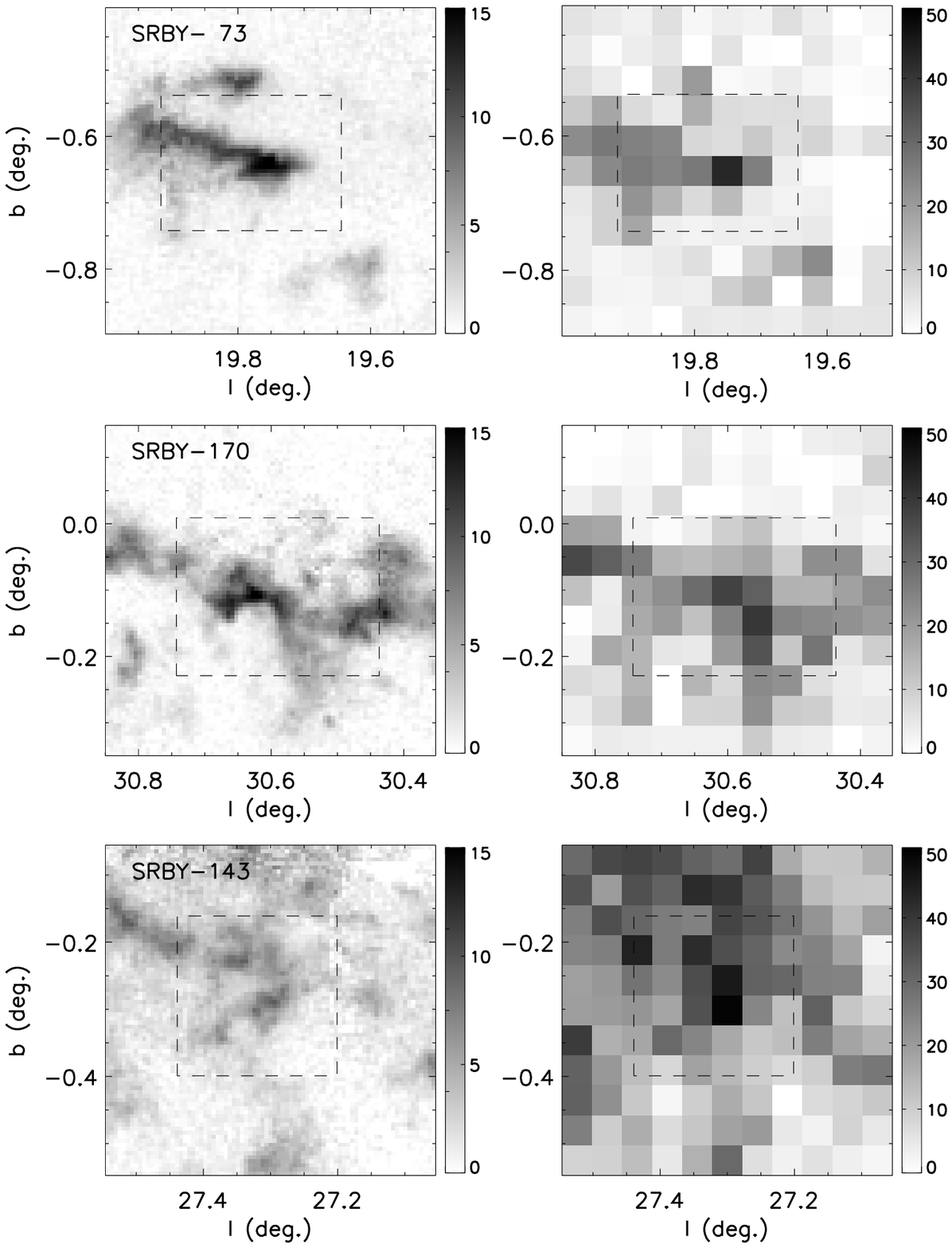}
\caption{(Left) Images of integrated \coa\ J=1-0 emission
 from the BU-FCRAO Galactic Ring Survey
and (Right) \co\ J=1-0 from the Massachusetts-Stony Brook Survey for 
3 giant molecular clouds catalogued by Solomon et~al. (1987) --
(Top) SRBY-73, (Middle) SRBY-170, and (Bottom) SRBY-143. 
The densely sampled, lower opacity \coa\ line emission offers a more 
detailed view of cloud structure than is revealed in the undersampled, 
\co\ data.  The dotted lines show the boundary of each cloud based on the 
emission centroid and angular sizes given by Solomon et~al. (1987).  
}
\label{fig1}
\end{center}
\end{figure}

The Boston University-FCRAO Galactic Ring Survey (GRS) 
imaged the \coa\ J=1-0 emission 
between Galactic longitudes 18$^\circ$ and 56$^\circ$ and latitudes,
$|b| \le 1^\circ$ with the FCRAO 14m telescope
(Jackson et~al. 2006).  
The advantages 
of the GRS over the UMSB survey includes higher angular sampling  
and spectral resolution and the use of a mostly optically  
thin tracer of molecular gas.  The lower opacity of \coa\ 
reduces, but does not fully eliminate, the effect of velocity crowding.
It also enables a more direct measure of molecular hydrogen 
column density and mass under the  assumption of 
local thermodynamic equilibrium (LTE) (Dickman 1978). 
These advantages are illustrated in Figure~\ref{fig1}, which 
compares the integrated emission from several clouds identified 
by SRBY 
derived from each survey.  
The UMSB \co\ data
identifies the 
location, angular extent and 
velocity 
of the molecular 
cloud but is unable to discern any substructure within the cloud. 
Owing to improved 
resolution afforded by fully sampling optically 
thin \coa\ emission, 
 the GRS data provide a more precise distribution 
of molecular material {\it within} the field  with reduced
confusion of signal from unrelated clouds along the line of sight. 
In this study, 
we examine the properties of the SRBY sample of GMCs
using the \coa\ J=1-0 data of the GRS.  

\section{Results}
For each GMC entry, 
the SRBY catalogue provides the emission weighted centroid 
positions, ($l_p$, $b_p$, $v_p$), angular extents along the galactic longitude and 
latitude axes, $\sigma_l, \sigma_b$, velocity dispersion, $\sigma_v$, \co\ luminosity,
$L_{CO,SRBY}$, virial mass, M$_{v,SRBY}$, and near/far side resolved distances, $D$.
Of the 273 GMCs catalogued by SRBY, 180 fall within the 
coverage of the GRS.  
All GMC kinematic distances are re-derived using the rotation curve of Clemens (1985).  However, 
random motions of clouds with respect to the velocity of the local standard of rest (LSR)
and streaming motions owing to spiral arm or localized perturbations introduce errors in 
such kinematic distances.  The fractional error, $\sigma_D/D$, can be large 
for nearby clouds. Over longitudes 20 to 50 degrees and \vlsr $<$ 20 \kms,  this fractional error ranges from 
50\% to 200\%  for velocity dispersions of 10 \kms.   Such errors propagate into significant uncertainties for the 
derived masses and sizes of the clouds.  Therefore, to minimize the 
fractional error when using kinematic distances, we restrict our analysis 
to 162 of these GMCs with \vlsr\ $>$ 20 \kms.  

A primary goal in this study is to derive the mass and surface density 
for each cloud over the same area as SRBY. 
We
consider the area, $A_1$, defined 
by the position centroids, $l_p,b_p$, and the extents, $\sigma_l, \sigma_b$,  for each 
angular axis as listed in the SRBY catalog 
\begin{equation}
A_1 = \int_{b_p-1.7\sigma_b}^{b_p+1.7\sigma_b}db \int_{l_p-1.7\sigma_l}^{l_p+1.7\sigma_l} dl \;\;\; deg^2
\end{equation}
where the factor, 1.7, comes from the relationship between 
the area of each cloud and $\sigma_l, \sigma_b$ described by SRBY.
The velocity interval for each cloud is determined from the 
inspection of the mean \coa\ and \co\ spectra over the area, $A_1$,
\begin{equation}
<T(v)> = { {\int_{A_1} dA\; T(l,b,v)} \over {\int_{A_1} dA} } \;\;\; K
\end{equation}
We also derive properties  within a secondary area, $A_2$, defined 
as the area within the half power isophote of the peak 
column density value within the cloud.  Typically, this area 
is 2-4\% of the SRBY defined area, $A_1$, and corresponds to the 
high column density central core of the cloud. 

For each of the 162 clouds within the GRS field and 
velocity range, the basic properties are re-calculated using the 
\coa\ data, $T(l,b,v)$.  The updated properties include emission 
weighted centroid 
positions, $l_\circ, b_\circ, v_\circ$, and 
velocity dispersion, $\sigma_v$, derived from the second moment of 
the mean \coa\ spectrum of the cloud, 
\begin{equation}
 \sigma_v^2 = {{\sum <T(v)>(v-v_\circ)^2}\over{\sum <T(v)>}} \;\;\; km^2 s^{-2}  
\end{equation}
The improved angular and spectral 
sampling of the GRS data provide a more precise 
position and velocity of the cloud than those provided in 
the SRBY catalog values.  The recomputed values for this sample of 
clouds are listed in Table~1. 

\subsection{GMC Masses}
The availability of lower opacity \coa\ J=1-0 emission from the GRS 
affords a more direct calculation of molecular hydrogen column densities 
and masses than is provided by \co.  The \coa\ column density is derived assuming 
LTE 
conditions within the cloud volume.  For each line of sight, $l,b$,
\begin{equation}
 N_{13}(l,b)=2.6{\times}10^{14} \left({{\tau_\circ}\over{1-exp(-\tau_\circ)}}\right)
{{\int T(l,b,v)dv} \over {(1-exp(-5.3/T_x)})} \;\;\; cm^{-2} 
\end{equation}
where the line center opacity, 
\begin{equation}
\tau_\circ = -ln\left[1-{{T_B}\over{5.3}}\left( (exp(5.3/T_x)-1)^{-1} - 0.16 \right)^{-1} \right] 
\end{equation}
(Rohlfs \& Wilson 2003). 
The excitation temperature, $T_x$, at each position is determined from 
the peak temperature of the optically thick \co\ line of 
UMSB survey data resampled to the GRS grid 
over the same target velocity interval.  The distribution of all spatially resampled 
$T_x$ values within the area, $A_1$
is shown in Figure~2.  Typical temperatures of GMCs range from 10 to 30 K so the
derived excitation temperatures imply either sub-thermal excitation conditions for most lines of 
sight or a non-unity filling factor of \co\ emission within the 45\arcsec\ beam of the FCRAO 
telescope.  

To relate this column density to the more abundant \htwo\ component, one 
requires the abundance ratio of \co\ to \coa\ and the ratio
of \co\ to \htwo.  It has long been established that the $^{12}C$ to 
$^{13}C$ abundance systematically varies with galactocentric 
radius, $R_{gal}$ (Penzias 1980).  The most recent characterization of this
gradient 
by Milam et~al. (2005) is 
\begin{equation}
[^{12}C/^{13}C] = 6.2 R_{gal} + 18.7
\end{equation}
This scaling is applied to the \coa\ column density for 
each cloud according to its Galactocentric
radius, $R_{gal}$, to derive a \co\ column density,
$N_{12}(l,b)= [^{12}C/^{13}C] N_{13}(l,b)$.  We then derive 
an \htwo\ column density, $N_{H_2}(l,b)$, at each grid position, 
assuming a constant \htwo/\co\ abundance ratio
of 1.1$\times$10$^4$ (Freking, Langer, \& Wilson 1982). This value is derived from 
extinction measurements of the nearby Taurus and $\rho$~Oph clouds and is larger by factors of 2-3
than the abundances determined by Lacy et~al. (1994) for lines of sight in the NGC~2024
and NGC~2264 clouds.  The adopted abundance values do not account for variations within the cloud 
owing to photochemistry, fractionation in the outer envelopes, or depletion of carbon onto 
dust grains in the cold, high density cores of the cloud.  Under these assumptions, the \htwo\ mass, 
$M_{LTE}$, is calculated from integration 
of the column density distribution, $N_{H_2}(l,b)$ over the solid angle of each area, $A_1$ and $A_2$, 
\begin{equation}
 M_{LTE} = \mu m_{H_2} D^2 \int d\Omega\; N_{H_2}(l,b)
\end{equation}
where $m_{H_2}$ is the mass of molecular hydrogen, $\mu=1.36$, is the mean 
molecular weight that accounts for the contribution of Helium, and $D$ is the 
distance to the cloud.  These masses are listed in Table~1. 

A comparison 
of the virial mass, $M_{v,SRBY}$, listed in the SRBY catalog, with $M_{LTE}$ within $A_1$, is 
shown in Figure~3.  While the values are well correlated, the 
LTE-derived masses are 
significantly smaller than the SRBY \co\ virial mass over the same projected 
area.  Typically,
$M_{LTE}\approx M_{v,SRBY}/5$. One may expect differences in cloud masses  respectively 
derived from \co\ and \coa\ owing to the relative opacities of the two observed lines
and photochemistry 
so these may not probe equivalent cloud volumes.  However, the mass discrepancy illustrated in 
Figure~3 is larger than differences measured for clouds in the Solar neighborhood (Heyer et~al. 2006; 
Goldsmith et~al. 2008).  
In \S2.3, we discuss the errors and uncertainties 
 associated with each mass estimate.

\subsection{GMC Surface Densities}
The surface density of a molecular cloud is a key property to its  
evolution and dynamical state (McKee \& Ostriker 2007). 
GMC surface densities are simply the mass of the cloud divided by 
the projected area.  Since we are tabulating masses within the same 
area as SRBY, the discrepancies in masses, discussed in the 
previous section, are mirrored in the resultant mass surface densities. 
The distribution of LTE derived surface 
densities determined within the SRBY defined areas is shown in Figure~4. 
Also shown is the distribution of mass surface density values within the 
half-power isophote of column density, $A_2$.  The vertical dotted line shows
the median surface density determined by SRBY. 
 For area $A_2$ that tracks the 
highest column density zones within each GMC, the LTE derived surface densities 
are comparable to the SRBY values that consider the entire GMC. 
The median surface density implied by the LTE derived masses within $A_1$
is 42 \mpcsq\ with a standard deviation of 37 \mpcsq. 
There are 
several reasons to expect a limited variation of GMC surface densities 
derived from a given gas tracer. 
First, for a given UV radiation field, there is a minimum column density 
necessary to self-shield \htwo\ and \co\ in order to build and 
maintain 
significant molecular abundances.  
Secondly, high density 
regions (cores) within 
a GMC subtend a small fraction of the projected area of a cloud and 
do not significantly contribute to the overall mass.  Moreover, 
owing to high optical depths and chemical depletion, 
such regions are not readily 
detected by \co\ or \coa. Therefore, one expects to find a limited 
range of molecular surface densities corresponding to those required for 
molecular self-shielding within a given UV radiation field (Elmegreen 1989; McKee 1989). 

\subsection{Errors and Uncertainties in \htwo\ Mass Determinations}
The large differences in mass and surface density of GMCs derived by SRBY and this study
suggest that one or both methods are subject to errors.  The differences are not due to 
instrumental errors of the CO measurements.  Masses are integrated quantities so the statistical 
errors are typically small (1-10\%).  However, both methods are affected by systematic errors, 
which we now summarize. 

The SRBY estimates of cloud virial 
masses rely on the accurate extrapolation of 
cloud properties, $\sigma_v,R$, from the values defined at the 
4 K or higher threshold of 
antenna temperature to the 1 K isophote that presumably circumscribes 
the bulk of the GMC.  
For such an extrapolation to succeed, the measured 
variation of the cloud properties (CO luminosity, size, velocity dispersion) with 
antenna temperature above the 4 K threshold must accurately reflect the 
structure of the cloud at all antenna temperature values.  However, the 
profile of these values above the 4 K threshold 
is limited by the angular undersampling of the UMSB survey and the 
opacity of the CO 
J=1-0 line.  Figure~1 demonstrates that the undersampled UMSB data 
misses much of the underlying 
structure of the cloud, especially within the 
brightest sub-regions, where the cloud is presumably defined at the 4 K limit
by SRBY.  
The primary effect of undersampling  is aliasing of small scale 
structure to larger scales. This aliasing  affects the 
inferred variation of cloud properties with antenna temperature.  For example, 
the 3\arcmin\ sampling with 45\arcsec\ resolution is unlikely 
to accurately measure the position and amplitude of localized  
emission peaks or troughs within the cloud yet must assign any detected signal to a
solid angle defined by the sampling interval.  
By not accounting for 
small scale structure, the derived variations of velocity dispersion, size, and luminosity 
with antenna temperature are less reliable.  
In addition, \co\ J=1-0 emission 
is strongly saturated owing to high optical depth.  Not only does high 
opacity 
obscure underlying cloud structure, but line saturation also 
flattens the surface brightness profile.
The corresponding extrapolation of shallow profiles induced by 
high optical depth 
of the \co\ line leads to overestimates 
of cloud sizes and CO luminosities at the 1 K isophote. 

A comparison of the extrapolated values of the CO 
luminosity, $L_{CO,SRBY}$, as listed in the catalog, and a direct 
measure, $L_{CO}$, determined from the integration of the UMSB CO intensities 
within the cloud boundaries and velocity intervals, is shown in Figure~5.  The extrapolated 
values are typically 35\% larger than the direct measures of $L_{CO}$.  Such 
higher values are 
unexpected given that the direct measure should be contaminated by signal 
from unrelated clouds along the line of sight over the same velocity range. 
This suggest that 
the extrapolated values of the CO luminosity are systematically 
overestimating the 
true CO luminosity. Similarly, the extrapolated 
cloud sizes and velocity dispersions that are used to derive the virial 
mass may also be inappropriate.  

Masses and \htwo\ column densities 
assuming LTE are also subject to several sources of systematic error. 
First, the assumption of equal \co\ and
\coa\ excitation temperatures is not valid for the full range of volume densities 
within molecular clouds.  Such excitation differences arise from 
the relative optical depths of the two lines such that these may not sample the 
same volumes and conditions within the cloud.  For densities less than the CO 
critical density ($n(H_2) <$ 750 \cc), the \co\ J=1-0 line can be thermalized by 
radiative trapping while the \coa\ J=1-0 line remains subthermally excited. 
Overestimating the \coa\ excitation temperature leads to an underestimation
of the \coa\ J=1-0 opacity.  
Second, the excitation of the rotational energy levels of \coa\ 
are not fully thermalized for most of the cloud volume.  
For densities less than 5$\times$10$^5$ \cc, the approximation of the partition function,
$\sum_{J=0}^{\infty}(2J+1)exp(-J(J+1)hB/kT_x) = kT/hB$, to account for 
material within the upper excitation states, overcorrects for the 
upper population energy levels. 

To examine these effects more quantitatively, 
we generate model \co\ and \coa\ line intensities for several sets of cloud conditions
using a large velocity gradient (LVG) approximation to account for non-thermal excitation.
The input model parameters are kinetic temperature, \htwo\ column density, $N_{model}$, \htwo\ 
volume density, $n(H_2)$, 
\co\ and \coa\ abundance, and velocity dispersion.  The filling factor of both \co\ and \coa\ 
emission is assumed to be unity.  For each model, the output 
\co\ and \coa\ line intensities are used to derive an LTE \htwo\ column density, $N_{LTE}$, 
that can be compared to the input value.  Figure~6 shows the fraction of
column density recovered by the LTE method as a function of volume density for models with 
kinetic temperatures of 8 and 15 K, \htwo\ to \co\ abundance of 1.1$\times$10$^4$, a 
\co\ to \coa\ abundance of 50, velocity dispersion of 2 \kms, for two values of $N_{model}$.  
In the low density regime ($n(H_2) <$ 400 \cc),
the models illustrate that  the LTE method underestimates the input column 
density by factors of $\sim$2 owing to an inappropriately high excitation temperature estimated 
from the \co\ line intensity
that leads to a reduced value of the \coa\ opacity.  This effect is more pronounced for the models 
with 8 K kinetic temperature.  With increasing density, the \coa\ J=1-0 
line becomes progressively thermalized such that the \coa\ opacity is more accurately determined but the upper rotational energy levels remain subthermally excited.  In this regime, LTE overestimates the column density by 10-40\%.  A similar result is obtained by Pineda et~al. (2008) and Goodman et~al. (2008), who used a simple  
curve of growth analysis to demonstrate that the \coa\ integrated emission increases more rapidly with extinction
in the low column density regime.  If these intensities are multiplied by a conversion factor 
derived from a linear fit over the full range of extinction, then the resultant 
column densities may overestimate the true values. 
In the high density 
limit ($n(H_2) >$2$\times$10$^5$ \cc), the assumption of LTE is valid such that 
$N_{LTE}/N_{model}$=1.  The models illustrate that column densities derived from \co\ and \coa\ J=1-0 
measurements assuming LTE can both underestimate or overestimate the true column densities depending 
on the physical conditions of the cloud.  

The assumption of constant abundance of CO relative to molecular hydrogen within a 
cloud is another source of of error in our application 
of the LTE method.  CO abundances can strongly vary 
between the strongly self-shielded interiors and the UV exposed 
envelope 
owing to selective photodissociation and fractionation (van Dishoeck \& Black 
1988; Liszt 2007). 
The effect of reduced CO to \htwo\ abundances in the cloud envelope 
is to mask regions that contain high column densities of \htwo\ but radiate no 
detectable \coa\ J=1-0 emission.  For example, the solid curve in Figure~6
corresponding to $N_{model}$ = 2$\times$10$^{21}$ cm$^{-2}$ 
is degenerate with an \htwo\ column density 
of 2$\times$10$^{22}$ cm$^{-2}$ and an \htwo\ to \co\ abundance of 1.1$\times$10$^5$. For densities 
less than 500 \cc, the predicted \coa\ emission is less than 0.55 K and would not be detected by the GRS 
at a 3$\sigma$ confidence level. 
By not considering 
these abundance variations  our 
LTE derived GMC masses may underestimate the true values.  There is insufficient 
information with these data to uniquely quantify this error. In a recent study of the 
Taurus molecular cloud, Goldsmith et~al. (2008) show that 
this subthermally excited, UV exposed envelope with no detectable \coa\ emission 
can contribute as much as 50\% of the cloud's mass when accounting for 
such molecular abundance variations.  
Based on this example and the fraction of area with detectable \coa\ emission 
within the SRBY boundaries, the true values of GMC mass and mass surface density could be 
larger by factors of 2-3 than the LTE derived values.  

A commonly used validation of the SRBY estimates of cloud masses and 
surface densities is the consistency with the CO intensity to \htwo\ conversion 
factor, $X_{CO}$, derived from $\gamma$-ray measurements (Bloemen 1986).
In fact, their implied value of $X_{CO}=M_{v,SRBY}/L_{CO,SRBY}$ varies by a factor of 2 over the luminosity range 
of the sample owing to the 4/5 power relationships dependence of $M_{v,SRBY}$ on $L_{CO,SRBY}$.
 For the median SRBY luminosity of 4.3$\times$10$^4$ $Kkms^{-1}pc^2$, the SRBY conversion 
factor is 5.7 $M_\odot/(Kkms^{-1} pc^2)$ or equivalently, 2.7$\times$10$^{20}$ H$_2$ molecules cm$^{-2}/(Kkms^{-1})$,
accounting for the contributions of Helium.
More recent estimates of $X_{CO}$ from $\gamma$-ray measurements indicate a lower value, 
$X_{CO} = 4 M_\odot/(Kkms^{-1}pc^2)$, which
corresponds to $1.9\times10^{20}$ H$_2$ molecules cm$^{-2}(Kkms^{-1})$ (Strong \& Mattox 1996).   
The implied 
SRBY \htwo\ column density is $9.7\times$10$^{21}$ cm$^{-2}$ corresponding to 
a \co\ surface 
brightness within the cloud boundaries of 51 K km s$^{-1}$ assuming this most recent 
CO to
\htwo\ conversion factor.
We have calculated the \co\ surface brightness for each cloud, $\int\int dv\;dA\; T(l,b,v)/\int dA$ using the 
UMSB data.  The 
median of the distribution of GMC surface brightness values is 30 K km s$^{-1}$.
This difference is evident for the 3 clouds shown 
in Figure~1, where the color lookup table for the \co\ images is set so all pixels 
above 50 K km s$^{-1}$ should be saturated.  Only a small fraction 
of pixels exceed this value. Cloud blending should only increase the 
mean surface brightness of each cloud as more of the SRBY defined area is filled by emission from 
unrelated clouds. 
Either the extrapolated 
SRBY column densities are too large or the \co-\htwo\ conversion factor 
is too small to be consistent with the observed surface brightness 
distributions. 

How do the newly derived LTE masses compare with the X factor? 
$M_{LTE}$ is linearly correlated with $L_{CO}$ in contrast to the 4/5 power relationships found by SRBY.  
The median value of $M_{LTE}/L_{CO}$ 
is 1.6 $M_\odot/(Kkms^{-1}pc^2)$. 
However,  $M_{LTE}$ is likely a lower limit due to the neglect of 
abundance variations within the outer envelope.  If $M_{GMC} \sim 2M_{LTE}$, as suggested 
by local cloud studies, 
then these rescaled LTE-derived column density estimates are compatible with the 
value of $X_{CO}$ determined from $\gamma$-rays.

\section{Discussion}
The GMC masses derived in this study are systematically lower 
than those estimated by SRBY. These lower values have significant 
implications to the global molecular content of the Galaxy. 
Solomon \& Rivolo (1989) provide a detailed summary of the biases
and completeness inherent in their cloud definition.  They 
demonstrate that the SRBY catalog is 
complete to a limiting mass of 
2.5$\times$10$^5$ \msun\ (corrected for current Galactic distances).  
Accounting for Malmquist bias effects,
they estimate that the SRBY clouds account for 40\% of the 
total CO luminosity and 40\% of the flux of the UMass-Stony Brook Survey. 
The remaining 
60\% of the CO luminosity and flux is presumed to originate from cold and/or 
small 
clouds that fall below their GMC identification threshold. 
The mass distribution, dN(M)/dM, of GMCs 
follows a power relationships,
$ dN/dM \sim M^{-\alpha_M} $
with $\alpha_M$ ranging from 1.5 (SRBY) to 1.8 (Heyer, Carpenter, \& Snell
2001).  The measured slopes of the GMC mass function imply that most 
of the molecular mass in the Galaxy resides 
within the largest clouds that are included in the SRBY sample. 
Yet, our new calculations of cloud masses imply that the 
true masses are smaller than the SRBY values by factors of 2-5 depending 
on the correction for subthermally excited gas and varying abundances within a 
cloud. Such a rescaling of GMC masses 
implies a reduced molecular mass content of the Galaxy by these same 
factors. 

Could there be a significant molecular gas mass component residing 
within smaller, cooler clouds that were not included in the SRBY 
catalog?  Owing to higher angular sampling than the UMSB survey, the GRS is more sensitive
to smaller clouds.
A comparison of the GRS field with SRBY 
clouds does indeed show both discrete and diffuse features 
that are not included within the SRBY catalog. Yet, we find that 32\% of the \coa\ emission 
over the GRS field originates within SRBY 
cloud boundaries, which is comparable to the value (40\%) estimated by 
Solomon \& Rivolo (1989) for \co.  The emission from volumes external to the SRBY boundaries 
may provide a significant and unaccounted reservoir of molecular material within the Galaxy.

The molecular gas fraction depends on both the surface density 
of the cloud and the ambient radiation field (Elmegreen 1989;
van Dishoeck \& Black 1988).  Significant abundances of 
\htwo\ and CO require sufficient column densities to self-shield.
It is possible that the fraction of LTE-derived mass to the 
SRBY-derived virial masses, $M_{LTE}/M_{v,SRBY}$,
and the higher SRBY surface densities reflect a
more intense ambient radiation field in the inner Galaxy owing to 
higher star formation activity
relative to the Solar neighborhood.  In this case, one would 
expect to find a dependence of this fraction
on Galactocentric radius.  However, we find no evidence for 
any variation of this fraction over the range of radii (4.1-8.1 kpc) 
of the cloud sample. 

\subsection{Re-examining Larson's Scaling Relationships}
The scaling relationships identified by Larson (1981) have provided 
a fundamental, observational constraint to descriptions 
of cloud dynamics and star formation.  The study by SRBY seemingly 
confirmed these scaling relationships for a larger sample of molecular 
clouds distributed throughout the Galaxy.  Given the results in the 
previous section, which demonstrate that the SRBY GMC masses 
and surface densities 
are likely overestimates to the true values, it is useful to re-examine 
the Larson (1981) scaling relationships with the new GRS data within the 
SRBY defined cloud boundaries. 

As has been demonstrated by many previous 
studies, 
the Larson scaling relationships are algebraically  
linked.  Here, we 
resummarize this coupling to
derive the coefficients that are critical to the interpretation 
of cloud dynamics.
Gravitational equilibrium (Larson's second relationship) 
implies that the observed mass of the cloud,
$M_{obs}$, is equal to 
the virial mass,
\begin{equation}
M_{obs} = 5\sigma_v^2R/G
\end{equation}
The cloud size, R, is defined as the radius of a circle with the equivalent area of the cloud.
The molecular gas surface density, $\Sigma$, of a cloud is simply the 
\htwo\ mass divided by the projected area,
\begin{equation}
 \Sigma = {{M_{obs}}\over{\pi R^2}} 
\end{equation}
Eliminating $M_{obs}$ and solving for $\sigma_v$,
\begin{equation}
 \sigma_v = (\pi G/5)^{1/2} \Sigma^{1/2} R^{1/2} 
\end{equation}
If $\Sigma$ is approximately constant for all clouds (Larson's third 
relationship), 
then one recovers the size-line width scaling (Larson's first relationship),
$\sigma_v=v_{\circ,G} R^{1/2}$, 
with the normalization coefficient 
\begin{equation}
 v_{\circ,G}=(\pi G \Sigma/5)^{1/2} 
\end{equation}
\begin{equation}
v_{\circ,G} = 0.52 \Bigl( { {\Sigma} \over {10^2\; M_\odot pc^{-2}}}\Bigr)^{1/2} \;\; km s^{-1} pc^{-1/2}
\end{equation}
The coefficient, $v_{\circ,G}$, parameterizes the scaling of velocities within a cloud such that 
non-thermal pressure balances the self-gravity of the cloud with surface density, $\Sigma$, and radius, $R$.

More generally, the velocity field of an interstellar cloud is described by the structure function that 
measures the variation of velocity differences of order p, with spatial scale, $\tau$, 
\begin{equation}
S_p(\tau) = \langle |v(\mathbf{x})-v(\mathbf{x}+\mathbf{\tau})|^p\rangle
\end{equation}
where the angle brackets denote a spatial average over the observed field.  
Within the inertial range, the structure function is expected to vary as a power relationships with $\tau$. For p=1,
\begin{equation}
S_1(\tau) = {\delta}v = v_\circ \tau^\gamma 
\end{equation}
where $\gamma$ is the scaling exponent and $v_\circ$ is the scaling coefficient.  These parameters correspond to 
Type 4 size-line width relationships described by Goodman et~al. (2004).  
The velocity dispersion
\footnote{Cloud-to-cloud size-velocity dispersion relationships 
use the full velocity dispersion of the cloud but scaled to the cloud radius, $R \sim L/2$.  Therefore, the 
respective definitions for the coefficient may differ by a factor of $\sim 2^\gamma$.}
 of an individual cloud is simply the structure function evaluated
at its cloud size, $L$, such that  $\sigma_v=S_1(L)=v_\circ L^\gamma$.  
Cloud-to-cloud size-velocity dispersion relationships, defined as Type 2 by Goodman et~al. (1998), 
are constructed from the end-points of each cloud's velocity structure function. 
The existence of a cloud-to-cloud size-velocity dispersion relationship  
identified by SRBY necessarily implies narrow distributions
of the scaling exponent and coefficient respectively 
for all clouds (Heyer \& Brunt 2004).  Large variations of $v_\circ$ and $\gamma$ between clouds would induce 
a large scatter of points that is inconsistent with the observations. 
From Monte Carlo modeling of the scatter of the SRBY size-velocity dispersion relationship, Heyer \& Brunt (2004) constrained the variation of $\gamma$ and $v_\circ$ between clouds to be less than 20\% about the mean values that is indicative of a universal structure function.  This universality is also reflected 
in the structure functions of individual clouds as derived by Brunt (2003) 
and Heyer \& Brunt (2004) using Principal Component Analysis.   

The Larson scaling relationships 
are concisely represented within the plane defined by the gas surface density, $\Sigma$, and the 
quantity, $\sigma_v/R^{1/2}$, for a set of GMC properties (see equation 10).  This representation assumes a scaling exponent 
of 1/2 for the structure function of each cloud so that the ordinate, $\sigma_v/R^{1/2}$, is 
equivalent
to the scaling coefficient, $v_\circ$. 
Absolute adherence 
to universality and all three of Larson's scaling relationships for a 
set of clouds 
would be ideally represented 
by a single point
centered at 
$\sigma_v/R^{1/2} = ({\pi}G\Sigma/5)^{1/2}$ for a 
constant value of $\Sigma$.
Given uncertainties in distance and deriving 
surface densities, one more realistically expects a cluster 
of points at this location. 
In Figure~7, 
we show the corresponding points derived 
from the GRS data within the SRBY boundaries (area $A_1$) and 
the area within the half-power isophote of $N_{H_2}$ (area $A_2$).
The vertical error bars displayed in the legend 
reflect a 20\% uncertainty in the distance to each cloud.
As a reference point, the large triangle denotes the location
of the SRBY median values ($\sigma_v/R^{1/2} = 0.72$ \kms; $\Sigma(H_2) = 206$ 
\mpcsq).  
The solid line shows the loci of points assuming gravitationally 
bound clouds,
$\sigma_v/R^{1/2}=(\pi G/5)^{1/2} \Sigma^{1/2}$ that is nearly 
identical to the coefficients used by SRBY.  
For both considered cloud areas, the \coa\ data points are 
displaced from this loci of virial equilibrium.
The median virial parameter, $\alpha_G=M_{v,13}/M_{LTE}$, is 1.9, where 
$M_{v,13}$ is the virial mass derived from \coa\ data within $A_1$. 
However, the LTE-derived mass could underestimate the true cloud mass by 
factors of 2-3 as suggested in \S2, so the derived properties are 
consistent with a 
virial parameter of unity for this sample of clouds. 

Figure~7 reveals a systematic 
variation of 
$v_\circ=\sigma_v/R^{1/2}$ with $\Sigma$. 
This trend is separately evident for each area,
$A_1$ (open circles) and $A_2$ (filled circles) with Pearson correlation coefficients 0.48 and 0.65 respectively.  For these sample sizes, it is
improbable that these data sets are drawn from a random population. 
The dependence of $\sigma_v/R^{1/2}$ on $\Sigma$ signals a 
departure 
from the universality of the velocity structure function of clouds.  It implies 
a necessary modification to Larson's scaling relationships 
but one that is 
compatible with the rather basic premise of 
gravitational equilibrium as described in 
equation 10.  
The measured variation of $v_\circ=\sigma_v/R^{1/2}$ 
is larger than the values derived by Heyer \& Brunt (2004) 
owing to the larger intrinsic scatter in the size-velocity dispersion relationship determined from the 
GRS data.   
 
The dependence of $\sigma_v/R^{1/2}$ on $\Sigma$ may not have been recognized in 
previous studies owing to a limited 
range of surface densities in the observed samples, or the use of a less reliable 
tracer of molecular gas column density, or simply not considered given the 
long-standing acceptance of Larson's scaling relationships.  The fidelity of the GRS data provides 
an excellent relative, if not absolute, measure of gas surface density that allows 
this relationship to be recognized.  We note that this relationship is algebraically imposed when 
deriving surface densities from the 
virial mass, $\Sigma = M_{vir}/\pi R^2 \propto \sigma_v^2/R$, as calculated by SRBY. 
However, as shown in Figure~8, the relationship is even evident in the SRBY defined properties
when using the mean \co\ surface brightness and CO to H$_2$ conversion factor as a measure of 
gas surface density, $\Sigma=X_{CO}L_{CO,SRBY}/\Omega_1 D^2$, where $\Omega_1$ is the 
solid angle of the cloud corresponding to $A_1$ and $D$ is the distance.   Moreover, the scaling 
between $\sigma_v/R^{1/2}$ and $\Sigma$ is 
also present in the sample of extragalactic GMCs tabulated by Bolatto et~al. (2008) (filled squares
in Figure~8).  The presence of this scaling within these independent data sets offers a powerful 
 verification that the velocity dispersion of a cloud depends on both the spatial scale of the emitting 
area and the mass surface density.  

\subsection{GMC Dynamics}
Descriptions of cloud dynamics
must consider the nature and origin of the 
observed supersonic motions in GMCs.  While much of the 
theoretical effort has focused on the scaling exponent of the 
power spectrum or structure function of the velocity field,  
the normalization, $v_\circ$, 
provides an important measure of the amplitude of these motions
as it is evaluated at a fixed scale of 1 pc.  Figure~7 
illustrates an additional constraint 
to these descriptions  -- that for a given cloud, 
the amplitude of the motions depends on the 
mass surface density.  It is not evident from the present measurements 
whether this variation of $v_\circ$ with $\Sigma$ is due to varying evolutionary 
states of the sample clouds or one that reflects different 
cloud conditions owing to the environmental diversity of the ISM. 

For sub-Alfv\'enic clouds whose neutral gas component is 
dynamically coupled to the 
interstellar magnetic field through ion-neutral collisions, the 
observed motions could arise from the propagation of large amplitude,
long wavelength Alfv\'en waves through the cloud 
(Arons \& Max 1975). 
In fact, a simple model of magnetically 
supported clouds in which the cloud surface density equals the 
magnetic critical surface density, $\Sigma_c=B/(63G)^{1/2}$, predicts the trend 
observed in Figure~7 (Mouschovias
1987; Myers \& Goodman 1988a; Mouschovias \& Psaltis 1995; Mouschovias, Tassis, \& Kunz 2006). 
The observed variation of the coefficient with surface density simply reflects 
plausible 
differences of the magnetic field strength 
between clouds owing to Galactic environments and 
the support of the cloud by the magnetic field such that $\Sigma \approx \Sigma_c$.
The vertical scatter of values for a given surface density would arise from varying 
flux-to-mass ratios owing to ambipolar diffusion. 

A definitive test of the role of the interstellar magnetic field in the support 
of the cloud and the origin of observed
internal motions requires measures of the magnetic field strength.  Such measurements are 
not available for this set of GMCs.   Myers \& Goodman (1988b) demonstrate
that the magnetic field strength derived from thermal OH Zeeman measurements is comparable to 
the predicted field strength assuming equipartition between the magnetic, kinetic, and 
gravitational energies, $B_{eq}\approx(45/G)^{1/2}\sigma_v^2/R$ for a set of nearby clouds. 

\section{Summary}
Our re-examination of the properties of GMCs in the Milky Way have 
identified two new results that challenge the long-standing assumptions of 
cloud dynamics.\\
1. The mass surface density of GMCs is lower than previously estimated by SRBY.
Assuming a constant abundance of molecular hydrogen to CO within a 
cloud, the 
median mass surface is 42 M$_\odot$pc$^{-2}$. 
Abundance variations within the outer envelope of clouds 
could increase the mass surface density to 80-120 M$_\odot$pc$^{-2}$.
No dependence of mass surface density is found with galactocentric 
radius. 
\\
2. The normalization of the velocity structure function, 
derived from the velocity dispersion and the size of each cloud, 
$v_\circ = \sigma_v/R^{1/2}$, varies with the molecular gas 
surface density, as $\Sigma^{1/2}$. The dependence of this factor 
on surface density conflicts with Larson's velocity scaling relationship
and the universality of turbulence within molecular clouds. 
However, this dependence is consistent with the prediction 
of Mouschovias (1987) that attributes the observed motions to 
Alfv\'en waves and the support of GMCs by the interstellar magnetic field.

\acknowledgments
This work was supported by NSF grant AST 0540852 and 0838222 to the Five College
Radio Astronomy Observatory.  
The authors acknowledge valuable discussions with 
Telemachos Mouschovias and critical comments by the referee.  
We also thank Alberto Bolatto for providing the table of extragalactic GMC properties prior to publication. 
Finally, while this study challenges the results in SRBY, we wish to acknowledge 
the many insightful contributions to studies of the 
molecular interstellar medium and star formation made by Phil Solomon, who passed away in 2008.

\newpage

\begin{figure}
\begin{center}
\includegraphics{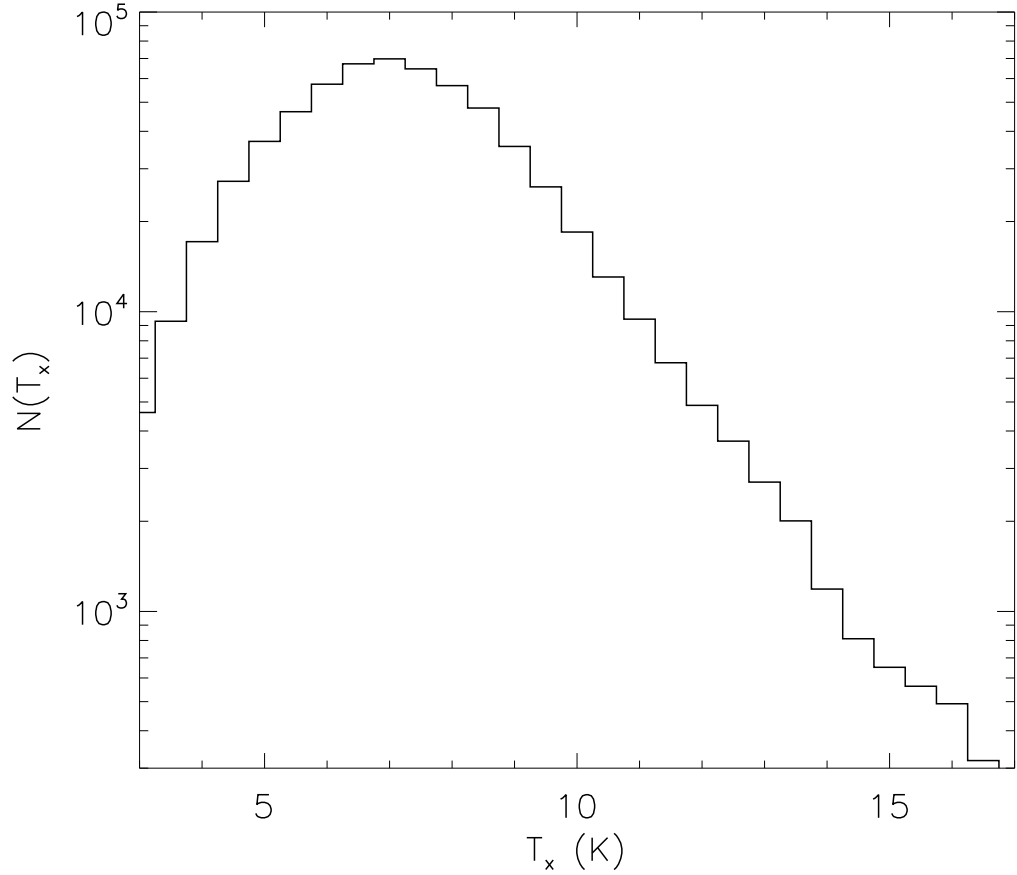}
\caption{The distribution of excitation temperature values within the SRBY defined areas 
derived by resampling the UMSB survey onto the GRS grid.  The excitation temperatures are smaller than kinetic 
temperature values expected for GMCs and denote cloud densities less than the 
critical density of the \co\ J=1-0 transition or a filling factor of \co\ emission less than unity 
over a 45\arcsec\ FWHM beam. }
\label{fig2}
\end{center}
\end{figure}

\begin{figure}
\begin{center}
\includegraphics{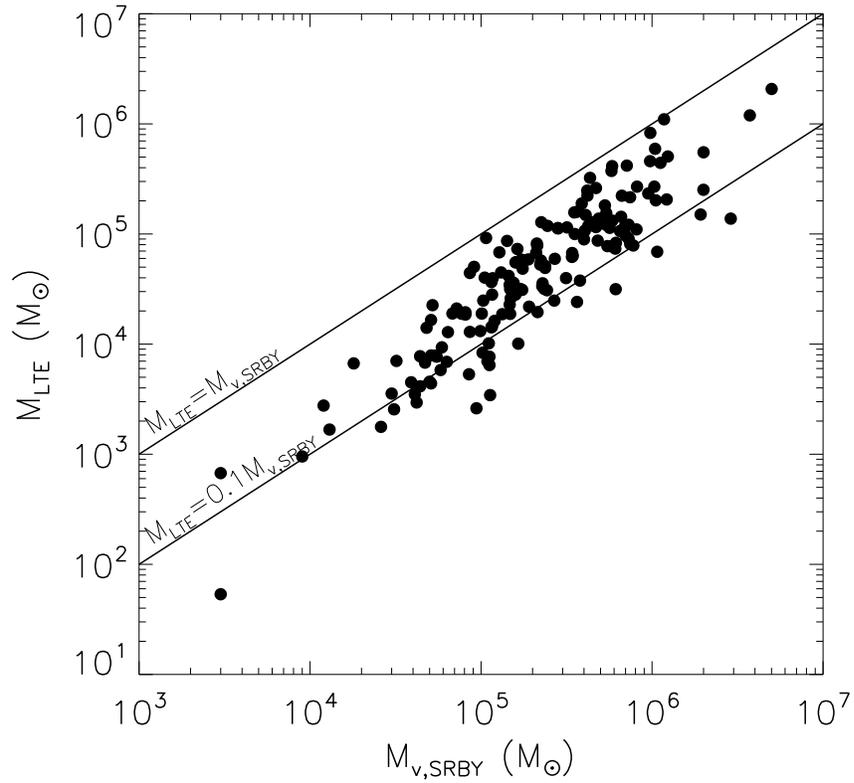}
\caption{A comparison of cloud masses enclosed within the SRBY defined areas 
calculated 
from \coa\ assuming LTE, $M_{LTE}$ with virial masses determined by SRBY, $M_{v,SRBY}$. 
The LTE masses are systematically lower than the SRBY derived 
virial masses.
}
\label{fig3}
\end{center}
\end{figure}

\begin{figure}
\begin{center}
\includegraphics{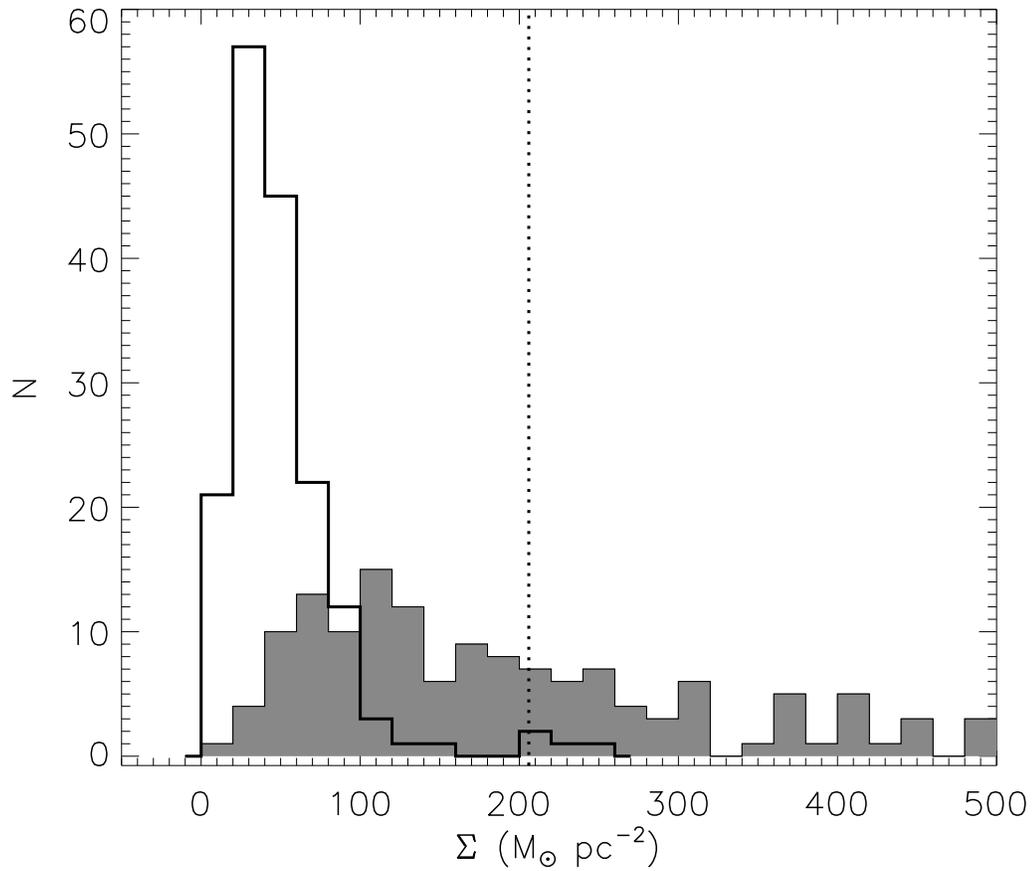}
\caption{The histogram of cloud mass surface densities derived from 
\coa\ data and assuming LTE and a constant \htwo\ to \co\ abundance ratio 
within areas, $A_1$ (heavy line) and $A_2$ (shaded).  The vertical dotted 
line denotes the median surface density from SRBY.  
}
\label{fig4}
\end{center}
\end{figure}

\begin{figure}
\begin{center}
\includegraphics{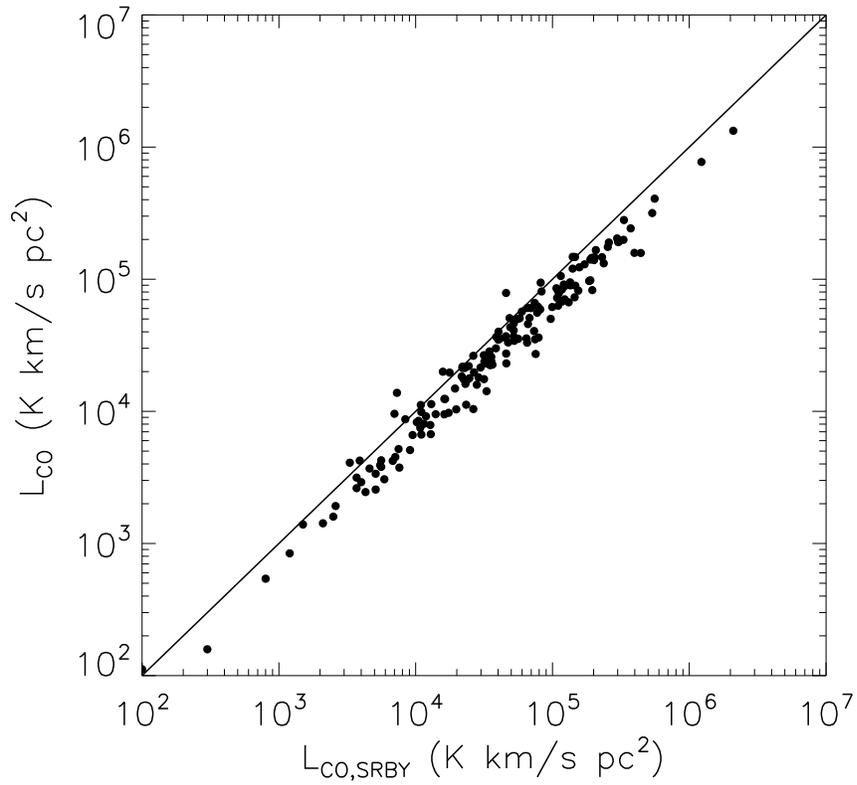}
\caption{$L_{CO}$ from the direct integration of the signal 
over the enclosed area versus the CO luminosity from the SRBY catalog. 
The solid line shows $L_{CO,SRBY}=L_{CO}$. 
The smaller direct integration values of $L_{CO}$ suggest that the extrapolated 
values overestimate the true value. 
}
\label{fig5}
\end{center}
\end{figure}

\begin{figure}
\begin{center}
\includegraphics{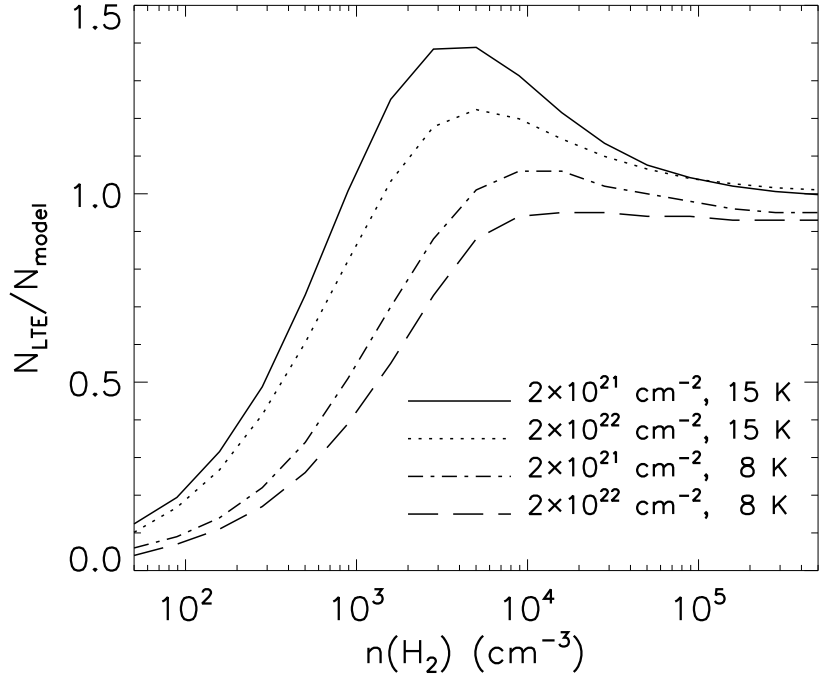}
\caption{The variation of the fraction of column density recovered by the LTE method, $N_{LTE}/N_{model}$,
with volume density as derived from LVG models with model column densities of 2$\times$10$^{21}$ cm$^{-2}$ 
and 2$\times$10$^{22}$ cm$^{-2}$, kinetic temperatures of 8K and 15 K, velocity dispersion of 2 \kms, and 
constant \co\ and \coa\ abundances. 
Over the range of densities expected for the bulk of molecular clouds (500 \cc $< n(H_2) <$ 5000 \cc), 
the LTE method can both underestimate and overestimate the gas column density owing to varying degrees
of excitation. 
}
\label{fig6}
\end{center}

\end{figure}

\begin{figure}
\begin{center}
\includegraphics{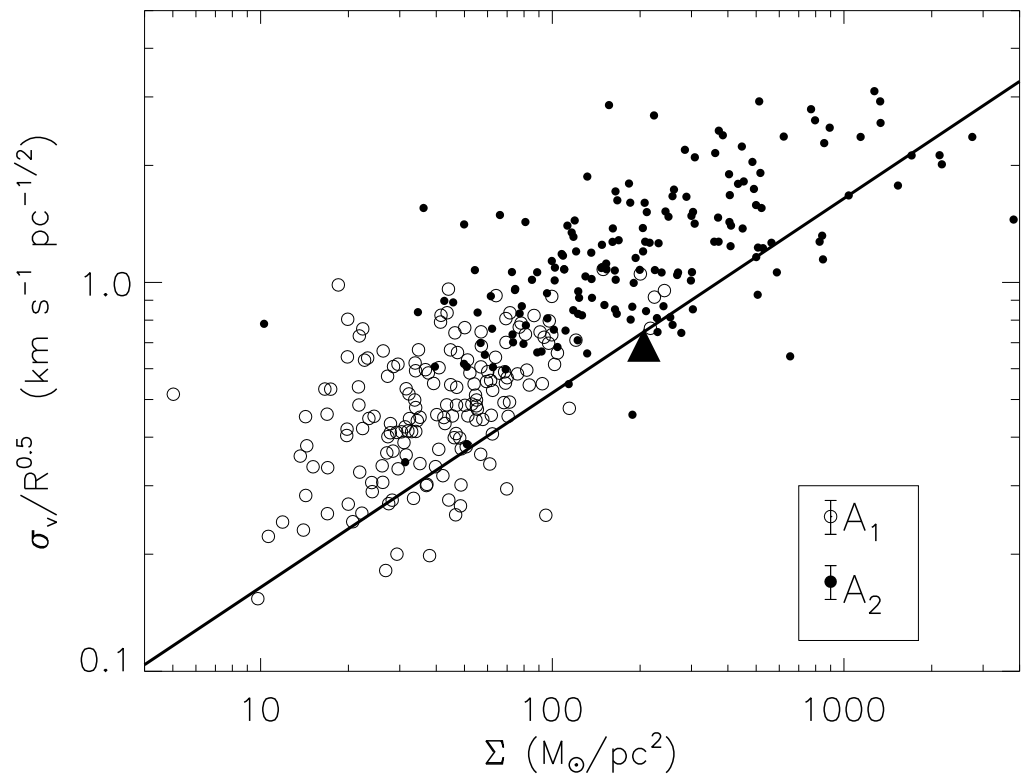}
\caption{The variation of the scaling coefficient, 
$v_\circ = \sigma_v/R^{1/2}$, with mass surface density
derived within the SRBY cloud boundaries (open circles) 
and the 1/2 maximum isophote of \htwo\ column density (filled circles).  
The filled triangle denotes the value derived by SRBY.
The solid line shows the loci of points corresponding to 
gravitationally bound clouds.  There is a dependence of 
the coefficient with mass surface density in contrast to 
Larson's velocity scaling relationship.  The error bars in the legend 
reflect a 20\% uncertainty of the distance to each cloud.
}
\label{fig7}
\end{center}
\end{figure}

\begin{figure}
\begin{center}
\includegraphics{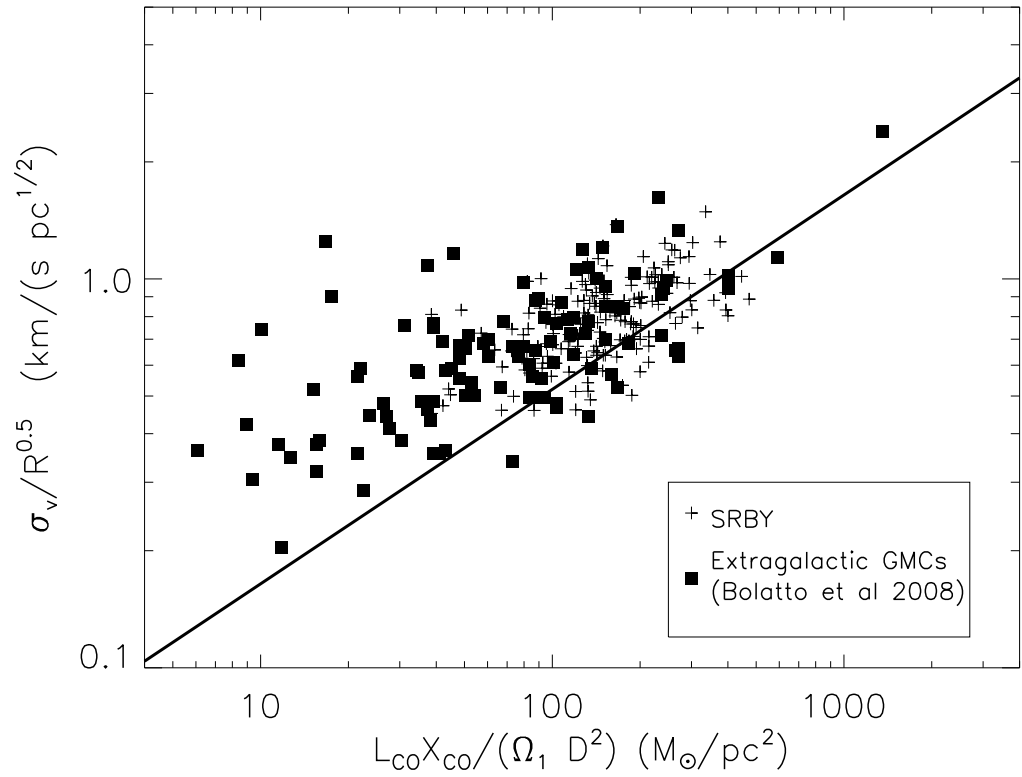}
\caption{The variation of the scaling coefficient, 
$v_\circ = \sigma_v/R^{1/2}$, with mass surface density for 
GMCs from the SRBY catalog (+ symbols) and extragalactic 
GMCs from Bolatto et~al. (2008) (filled squares).  
}
\label{fig8}
\end{center}
\end{figure}

\newpage
\begin{landscape}
\begin{deluxetable}{lcccccccccccccc}
\tablecolumns{15} 
\tablewidth{555pt}
\tabletypesize{\scriptsize}
\tablecaption{Rederived GMC Properties} 
\tablehead{
\colhead{}&\multicolumn{9}{c}{Within SRBY Defined Area, A$_1$} &\colhead{}& 
\multicolumn{4}{c}{Within Half Max Isophote of N(H$_2$), A$_2$} \\ 
\cline{2-10} \cline{12-15} \\ 
\colhead{SRBY} & \colhead{$l$} & \colhead{$b$} & \colhead{$v_\circ$} &  
\colhead{$R_g$} & \colhead{Dist} & \colhead{$\sigma_v$} & \colhead{R} & \colhead{$L_{CO}$} & \colhead{M$_{LTE}$} && \colhead{$\sigma_v$} & \colhead{R} & \colhead{$L_{CO}$} & \colhead{M$_{LTE}$}
\\
\colhead{} & \colhead{(deg)} & \colhead{(deg)} & \colhead{(km/s)} & \colhead{(kpc)} & \colhead{(kpc)} & \colhead{(km/s)} & \colhead{(pc)} & \colhead{($K km/s pc^2$)} & \colhead{(M$_\odot$)} & &
\colhead{(km/s)} & \colhead{(pc)} & \colhead{($K km/s pc^2$)} & \colhead{(M$_\odot$)}
}
\startdata 
64 &  18.87 &  -0.01 &   47.6 &  5.0 & 12.2 &   2.6 &  40.6 & 1.32e+05 & 3.23e+05 &&   2.0 &   5.5 & 3.46e+03 & 2.88e+04\\
 67 &  19.22 &  -0.23 &   37.2 &  6.0 & 13.3 &   1.5 &  42.2 & 8.27e+04 & 7.86e+04 &&   1.7 &   5.0 & 8.94e+02 & 4.88e+03\\
 68 &  19.31 &   0.03 &   26.0 &  6.5 &  2.2 &   1.9 &   6.9 & 3.91e+03 & 1.41e+04 &&   2.0 &   0.8 & 5.58e+01 & 8.98e+02\\
 70 &  19.61 &  -0.08 &   60.6 &  4.7 &  4.2 &   3.0 &   9.7 & 8.28e+03 & 1.31e+04 &&   2.6 &   2.3 & 4.83e+02 & 2.74e+03\\
 71 &  19.63 &  -0.66 &   56.7 &  4.9 &  4.1 &   2.2 &  10.3 & 8.47e+03 & 1.89e+04 &&   2.2 &   3.1 & 8.70e+02 & 4.47e+03\\
 72 &  19.67 &   0.11 &   25.0 &  6.5 & 13.8 &   2.4 &  41.3 & 1.23e+05 & 2.61e+05 &&   2.2 &   9.6 & 7.17e+03 & 3.54e+04\\
 73 &  19.78 &  -0.64 &   23.8 &  6.7 &  2.0 &   1.3 &   4.6 & 8.42e+02 & 2.77e+03 &&   0.9 &   0.5 & 7.73e+00 & 6.48e+02\\
 74 &  19.81 &  -0.39 &   68.2 &  4.3 &  4.7 &   2.7 &  32.1 & 6.30e+04 & 1.10e+05 &&   2.3 &   2.3 & 4.00e+02 & 3.50e+03\\
 75 &  19.90 &  -0.62 &   44.1 &  5.4 &  3.4 &   2.8 &  23.2 & 5.91e+04 & 1.28e+05 &&   2.3 &   1.6 & 3.39e+02 & 3.64e+03\\
 77 &  20.58 &  -0.42 &   63.3 &  4.5 & 11.3 &   2.9 &  33.8 & 6.20e+04 & 1.21e+05 &&   2.5 &   2.9 & 2.60e+02 & 9.78e+03\\
 78 &  20.69 &  -0.32 &   62.8 &  4.7 & 11.5 &   1.9 &  22.8 & 4.10e+04 & 7.85e+04 &&   1.7 &   5.2 & 3.41e+03 & 1.95e+04\\
 79 &  20.76 &  -0.09 &   57.8 &  4.8 & 11.7 &   3.1 &  27.1 & 9.12e+04 & 1.89e+05 &&   2.6 &   4.4 & 3.29e+03 & 2.48e+04\\
 80 &  20.75 &   0.05 &   78.2 &  4.2 &  5.0 &   3.1 &  16.7 & 1.75e+04 & 1.95e+04 &&   3.0 &   4.4 & 1.98e+03 & 4.92e+03\\
 81 &  20.87 &  -0.02 &   32.3 &  6.3 & 13.5 &   4.0 &  45.0 & 1.41e+05 & 2.33e+05 &&   3.7 &  11.7 & 1.34e+04 & 4.72e+04\\
 82 &  20.91 &  -0.31 &   66.3 &  4.5 &  4.6 &   1.4 &  15.3 & 1.61e+04 & 1.01e+04 &&   1.5 &   6.1 & 2.08e+03 & 4.63e+03\\
 84 &  21.54 &  -0.64 &   53.5 &  5.1 &  3.9 &   2.0 &  23.7 & 3.56e+04 & 4.92e+04 &&   1.6 &   0.3 & 9.61e+00 & 3.76e+02\\
 85 &  21.71 &  -0.02 &   68.3 &  4.6 &  4.5 &   1.9 &  18.1 & 2.31e+04 & 2.41e+04 &&   2.0 &   5.3 & 2.31e+03 & 6.94e+03\\
 86 &  21.36 &   0.00 &   74.4 &  4.3 & 10.9 &   1.7 &  32.0 & 7.85e+04 & 1.19e+05 &&   1.7 &   6.7 & 4.28e+03 & 1.86e+04\\
 87 &  21.52 &   0.26 &   78.8 &  4.3 &  5.0 &   2.2 &   7.5 & 4.24e+03 & 3.50e+03 &&   2.0 &   1.8 & 2.59e+02 & 6.73e+02\\
 88 &  21.87 &  -0.36 &   82.2 &  4.2 & 10.6 &   2.8 &  46.0 & 2.04e+05 & 2.02e+05 &&   2.5 &   6.0 & 3.29e+03 & 1.54e+04\\
 89 &  22.07 &   0.17 &   51.7 &  5.3 &  3.6 &   2.8 &  22.9 & 3.69e+04 & 6.19e+04 &&   1.8 &   2.0 & 4.19e+02 & 2.62e+03\\
 90 &  22.34 &   0.08 &   84.2 &  4.1 & 10.4 &   1.9 &  18.4 & 4.62e+04 & 5.83e+04 &&   1.8 &   2.0 & 3.84e+02 & 4.51e+03\\
 91 &  22.41 &   0.33 &   84.5 &  4.1 & 10.4 &   1.1 &  17.1 & 3.66e+04 & 4.43e+04 &&   1.2 &   6.9 & 3.92e+03 & 2.82e+04\\
 92 &  22.54 &  -0.04 &  115.1 &  3.3 &  7.9 &   0.5 &  10.6 & 1.03e+04 & 3.44e+03 &&   0.5 &   1.7 & 1.16e+02 & 4.63e+02\\
 93 &  22.56 &  -0.20 &   77.4 &  4.4 & 10.8 &   3.1 &  11.4 & 3.32e+04 & 9.20e+04 &&   3.1 &   8.8 & 1.27e+04 & 6.52e+04\\
 94 &  22.74 &  -0.24 &  106.3 &  3.4 &  8.7 &   2.8 &  16.0 & 2.26e+04 & 5.55e+04 &&   3.2 &   1.2 & 1.04e+02 & 2.32e+03\\
 95 &  22.81 &   0.41 &   91.9 &  3.9 & 10.0 &   2.3 &  19.8 & 4.58e+04 & 3.97e+04 &&   2.5 &  12.8 & 1.72e+04 & 2.92e+04\\
 96 &  22.86 &   0.40 &  114.1 &  3.3 &  7.8 &   2.0 &   6.4 & 9.59e+03 & 5.32e+03 &&   2.1 &   3.9 & 1.15e+03 & 3.47e+03\\
 97 &  23.00 &  -0.36 &   76.6 &  4.5 & 10.9 &   2.3 &  83.4 & 1.33e+06 & 2.08e+06 &&   2.2 &   5.6 & 6.26e+03 & 4.99e+04\\
 98 &  22.97 &  -0.02 &   80.2 &  4.3 & 10.6 &   1.6 &  28.2 & 1.47e+05 & 1.21e+05 &&   1.5 &   2.2 & 3.27e+02 & 4.54e+03\\
 99 &  23.07 &   0.64 &   37.2 &  6.1 &  2.8 &   0.8 &   8.8 & 1.60e+03 & 6.68e+03 &&   0.5 &   0.6 & 1.21e+01 & 7.40e+02\\
100 &  23.39 &  -0.23 &   99.6 &  3.7 &  9.3 &   4.6 &  23.3 & 9.70e+04 & 4.13e+05 &&   4.1 &   3.0 & 2.63e+03 & 3.22e+04\\
101 &  23.51 &  -0.40 &   74.9 &  4.6 &  4.7 &   1.5 &  11.6 & 1.99e+04 & 1.45e+04 &&   1.6 &   5.3 & 2.50e+03 & 7.03e+03\\
102 &  23.55 &   0.19 &   87.3 &  4.3 & 10.5 &   5.9 &  29.8 & 1.47e+05 & 4.18e+05 &&   5.8 &  12.2 & 3.84e+04 & 1.35e+05\\
103 &  23.68 &   0.52 &   83.2 &  4.3 &  5.1 &   2.0 &   9.7 & 4.22e+03 & 5.81e+03 &&   1.9 &   2.8 & 7.86e+02 & 2.45e+03\\
105 &  23.96 &   0.14 &   80.9 &  4.5 &  5.0 &   2.1 &   8.2 & 1.12e+04 & 2.10e+04 &&   2.2 &   2.2 & 6.33e+02 & 4.56e+03\\
106 &  24.21 &  -0.04 &   88.4 &  4.2 & 10.1 &   1.8 &  22.7 & 6.06e+04 & 8.19e+04 &&   2.1 &   3.8 & 1.77e+03 & 9.05e+03\\
107 &  24.45 &  -0.80 &   58.4 &  5.3 &  3.8 &   1.6 &   5.7 & 3.15e+03 & 4.55e+03 &&   1.9 &   3.9 & 1.35e+03 & 3.57e+03\\
109 &  24.49 &   0.20 &   36.4 &  6.2 & 12.8 &   1.7 &  25.4 & 3.54e+04 & 5.27e+04 &&   2.1 &   7.8 & 4.99e+03 & 2.12e+04\\
110 &  24.39 &  -0.24 &   56.1 &  5.1 & 11.5 &   2.7 &  19.3 & 3.62e+04 & 1.19e+05 &&   2.6 &  10.5 & 1.49e+04 & 6.43e+04\\
111 &  24.50 &  -0.15 &   44.9 &  5.9 &  3.1 &   1.7 &   9.7 & 1.24e+04 & 2.48e+04 &&   1.9 &   2.5 & 7.56e+02 & 4.02e+03\\
112 &  24.42 &  -0.41 &   44.3 &  5.9 &  3.1 &   0.9 &  20.2 & 2.72e+04 & 3.77e+04 &&   1.1 &   0.4 & 5.76e+00 & 2.47e+02\\
113 &  24.49 &  -0.72 &   48.6 &  5.7 &  3.2 &   1.0 &   9.1 & 5.09e+03 & 7.69e+03 &&   1.0 &   1.2 & 6.61e+01 & 6.19e+02\\
114 &  24.51 &  -0.23 &   96.6 &  3.9 &  9.3 &   2.5 &  26.4 & 6.83e+04 & 1.16e+05 &&   2.5 &   5.4 & 4.48e+03 & 2.06e+04\\
115 &  24.54 &  -0.50 &   60.6 &  5.1 &  4.0 &   1.9 &  12.0 & 1.78e+04 & 4.17e+04 &&   2.0 &   5.8 & 3.30e+03 & 1.76e+04\\
116 &  24.63 &  -0.14 &   83.8 &  4.4 & 10.3 &   2.1 &  25.8 & 8.20e+04 & 6.89e+04 &&   2.6 &   5.8 & 3.69e+03 & 1.16e+04\\
117 &  24.67 &  -0.05 &  110.3 &  3.5 &  7.7 &   3.0 &  26.3 & 3.70e+04 & 1.13e+05 &&   2.1 &   1.4 & 1.50e+02 & 9.43e+03\\
118 &  25.18 &   0.16 &  103.1 &  3.8 &  9.0 &   3.9 &  23.3 & 8.08e+04 & 1.19e+05 &&   3.4 &   4.5 & 2.60e+03 & 1.32e+04\\
119 &  25.27 &   0.33 &   45.1 &  5.8 & 12.2 &   1.8 &  34.7 & 1.06e+05 & 9.07e+04 &&   1.3 &   1.5 & 8.48e+01 & 4.16e+03\\
120 &  25.54 &  -0.39 &  116.3 &  3.7 &  7.7 &   1.8 &   7.3 & 3.75e+03 & 4.40e+03 &&   1.7 &   3.4 & 5.34e+02 & 2.24e+03\\
121 &  25.40 &  -0.24 &   58.2 &  5.0 & 11.2 &   4.1 &  52.0 & 1.58e+05 & 5.94e+05 &&   3.3 &   1.4 & 1.77e+02 & 4.75e+03\\
122 &  25.63 &  -0.11 &   94.2 &  4.2 &  9.7 &   2.5 &  53.7 & 4.07e+05 & 5.52e+05 &&   2.2 &   8.0 & 7.84e+03 & 5.20e+04\\
123 &  25.54 &  -0.21 &  118.1 &  3.7 &  7.7 &   1.7 &  16.3 & 1.12e+04 & 1.87e+04 &&   1.3 &   1.7 & 2.07e+02 & 1.73e+03\\
124 &  25.79 &   0.56 &   46.3 &  5.8 &  3.2 &   1.8 &  17.9 & 1.04e+04 & 3.16e+04 &&   1.8 &   6.0 & 2.22e+03 & 8.27e+03\\
125 &  25.72 &   0.24 &  110.7 &  3.8 &  8.5 &   2.0 &  42.7 & 1.92e+05 & 1.50e+05 &&   2.0 &   5.3 & 3.82e+03 & 2.12e+04\\
126 &  25.91 &   0.22 &   69.6 &  4.9 &  4.4 &   1.3 &   7.2 & 4.09e+03 & 3.55e+03 &&   1.5 &   2.8 & 3.33e+02 & 1.05e+03\\
127 &  25.71 &  -0.15 &  106.3 &  3.8 &  8.5 &   2.1 &  34.0 & 6.78e+04 & 1.15e+05 &&   2.1 &   3.7 & 1.26e+03 & 4.39e+03\\
128 &  25.90 &  -0.13 &  104.8 &  3.9 &  8.7 &   2.5 &  37.4 & 1.75e+05 & 2.06e+05 &&   3.1 &   7.0 & 4.33e+03 & 1.68e+04\\
129 &  25.96 &  -0.57 &   62.0 &  5.2 &  4.1 &   1.4 &   9.6 & 4.51e+03 & 4.14e+03 &&   1.5 &   3.2 & 4.26e+02 & 1.11e+03\\
130 &  26.18 &   0.13 &   70.5 &  4.9 &  4.5 &   1.5 &  12.8 & 7.88e+03 & 1.02e+04 &&   1.5 &   5.9 & 2.05e+03 & 5.45e+03\\
131 &  26.35 &   0.79 &   47.1 &  5.9 &  3.1 &   1.2 &   8.4 & 2.91e+03 & 7.04e+03 &&   1.7 &   0.4 & 3.28e+00 & 1.12e+02\\
133 &  26.55 &  -0.31 &  107.7 &  3.9 &  8.5 &   2.0 &  16.1 & 2.59e+04 & 4.48e+04 &&   1.9 &   1.8 & 2.82e+02 & 3.13e+03\\
134 &  26.60 &   0.01 &   26.8 &  7.0 & 13.4 &   2.5 &  25.8 & 9.47e+04 & 1.49e+05 &&   2.8 &   3.6 & 1.35e+03 & 1.02e+04\\
135 &  26.66 &   0.01 &   99.6 &  4.1 &  9.1 &   3.3 &  38.5 & 1.40e+05 & 8.11e+04 &&   3.0 &  11.4 & 1.38e+04 & 1.87e+04\\
136 &  26.68 &   0.52 &   87.2 &  4.5 &  5.2 &   1.6 &  12.2 & 6.62e+03 & 7.88e+03 &&   1.7 &   1.6 & 1.40e+02 & 9.37e+02\\
137 &  26.68 &   0.01 &  111.2 &  3.8 &  7.6 &   1.0 &  12.5 & 9.75e+03 & 6.97e+03 &&   1.1 &   3.3 & 6.31e+02 & 2.14e+03\\
138 &  26.93 &   0.13 &   92.4 &  4.4 &  5.5 &   1.7 &  16.9 & 2.38e+04 & 3.05e+04 &&   1.6 &   1.8 & 2.24e+02 & 1.38e+03\\
139 &  27.00 &  -0.39 &   68.1 &  5.0 &  4.3 &   1.8 &   6.1 & 2.45e+03 & 2.55e+03 &&   1.8 &   2.8 & 5.81e+02 & 1.34e+03\\
140 &  26.90 &  -0.11 &   79.8 &  4.7 &  5.0 &   1.6 &  17.7 & 9.50e+03 & 1.42e+04 &&   1.9 &   3.2 & 7.61e+02 & 2.85e+03\\
141 &  27.04 &  -0.15 &  102.8 &  4.1 &  9.0 &   1.4 &  14.8 & 2.18e+04 & 1.85e+04 &&   1.4 &   3.0 & 5.08e+02 & 2.72e+03\\
142 &  27.24 &   0.14 &   33.3 &  6.6 & 12.9 &   2.6 &  29.0 & 8.54e+04 & 1.32e+05 &&   2.4 &   5.1 & 2.83e+03 & 2.46e+04\\
143 &  27.32 &  -0.28 &   74.2 &  5.0 &  4.5 &   3.2 &  10.5 & 7.50e+03 & 6.45e+03 &&   2.6 &   3.4 & 8.48e+02 & 1.81e+03\\
144 &  27.35 &  -0.15 &   92.4 &  4.4 &  5.6 &   1.6 &   9.2 & 1.13e+04 & 1.66e+04 &&   1.5 &   1.1 & 1.26e+02 & 1.54e+03\\
145 &  27.52 &   0.21 &   35.4 &  6.5 & 12.7 &   1.0 &  17.0 & 2.46e+04 & 1.88e+04 &&   1.1 &   2.6 & 5.75e+02 & 2.21e+03\\
146 &  27.50 &   0.14 &   97.5 &  4.3 &  9.3 &   3.6 &  23.3 & 8.86e+04 & 1.55e+05 &&   4.1 &  10.4 & 1.78e+04 & 5.47e+04\\
147 &  27.63 &   0.10 &   82.9 &  4.7 &  5.1 &   1.5 &  15.0 & 1.71e+04 & 2.19e+04 &&   1.8 &   0.7 & 3.15e+01 & 5.57e+02\\
148 &  27.73 &   0.10 &   98.3 &  4.2 &  8.9 &   3.7 &  37.2 & 1.20e+05 & 2.48e+05 &&   3.4 &   9.9 & 1.29e+04 & 4.72e+04\\
149 &  28.19 &  -0.04 &   97.8 &  4.4 &  9.2 &   1.3 &  19.5 & 8.26e+04 & 8.32e+04 &&   1.5 &   3.4 & 1.27e+03 & 9.21e+03\\
150 &  28.24 &  -0.38 &   45.9 &  6.0 &  3.0 &   1.8 &  13.8 & 9.51e+03 & 2.60e+04 &&   2.0 &   0.7 & 3.54e+01 & 5.92e+02\\
151 &  28.32 &  -0.06 &   78.0 &  4.8 &  5.0 &   3.5 &  24.3 & 9.22e+04 & 2.23e+05 &&   3.3 &   4.7 & 4.07e+03 & 1.70e+04\\
152 &  28.61 &   0.05 &  101.3 &  4.3 &  8.8 &   4.1 &  26.5 & 1.48e+05 & 2.16e+05 &&   4.4 &   4.4 & 4.24e+03 & 1.87e+04\\
153 &  28.79 &   0.19 &   81.9 &  4.9 & 10.1 &   3.2 &  16.6 & 5.08e+04 & 6.74e+04 &&   3.1 &   8.1 & 8.35e+03 & 3.04e+04\\
154 &  28.80 &  -0.26 &   87.7 &  4.6 &  5.3 &   0.8 &  16.3 & 5.09e+04 & 3.15e+04 &&   1.1 &   2.2 & 5.95e+02 & 4.21e+03\\
155 &  28.98 &  -0.27 &   93.9 &  4.5 &  5.7 &   2.5 &  31.6 & 7.30e+04 & 1.82e+05 &&   3.5 &   3.4 & 1.94e+03 & 1.47e+04\\
156 &  28.99 &  -0.67 &   51.1 &  5.9 &  3.3 &   1.9 &   9.4 & 4.26e+03 & 9.35e+03 &&   1.6 &   1.6 & 1.66e+02 & 1.73e+03\\
157 &  29.35 &  -0.46 &   78.7 &  5.0 &  4.7 &   2.5 &  23.8 & 7.22e+04 & 9.68e+04 &&   2.8 &   1.3 & 4.46e+02 & 1.98e+03\\
158 &  29.01 &   0.05 &   96.9 &  4.4 &  8.9 &   2.9 &  41.1 & 1.90e+05 & 3.74e+05 &&   3.0 &   3.5 & 1.64e+03 & 7.12e+03\\
159 &  29.32 &  -0.57 &   63.9 &  5.4 &  4.0 &   1.7 &  14.0 & 1.49e+04 & 2.82e+04 &&   1.8 &   4.5 & 1.36e+03 & 7.51e+03\\
160 &  29.50 &   0.17 &   79.6 &  4.9 &  4.8 &   2.4 &  19.9 & 3.31e+04 & 5.87e+04 &&   2.4 &   8.5 & 9.00e+03 & 2.87e+04\\
161 &  29.61 &  -0.61 &   75.6 &  5.0 & 10.2 &   2.9 &  40.8 & 1.45e+05 & 2.69e+05 &&   2.0 &   6.1 & 2.54e+03 & 2.67e+04\\
162 &  29.89 &  -0.06 &   99.0 &  4.4 &  8.5 &   4.5 &  34.9 & 2.81e+05 & 8.28e+05 &&   4.1 &   6.0 & 1.14e+04 & 1.17e+05\\
163 &  29.90 &   0.10 &   39.3 &  6.4 & 12.2 &   2.1 &  12.9 & 1.75e+04 & 3.45e+04 &&   2.0 &   7.0 & 3.33e+03 & 1.56e+04\\
164 &  29.91 &  -0.77 &   83.7 &  4.8 &  5.1 &   1.1 &  13.2 & 6.72e+03 & 2.04e+04 &&   1.2 &   0.9 & 3.96e+01 & 1.44e+03\\
165 &  30.41 &   0.46 &   45.1 &  6.2 &  2.8 &   1.2 &   5.0 & 1.39e+03 & 1.67e+03 &&   1.4 &   1.4 & 1.45e+02 & 6.63e+02\\
167 &  30.56 &   0.32 &   92.2 &  4.5 &  8.7 &   1.9 &  26.1 & 9.42e+04 & 8.70e+04 &&   1.9 &   2.5 & 3.05e+02 & 2.37e+03\\
168 &  30.57 &  -0.02 &   40.8 &  6.3 & 11.9 &   2.9 &  44.7 & 1.47e+05 & 2.69e+05 &&   3.0 &   2.8 & 8.95e+02 & 1.07e+04\\
169 &  30.61 &  -0.45 &   94.0 &  4.6 &  8.8 &   2.4 &  14.4 & 4.99e+04 & 3.35e+04 &&   2.5 &   8.6 & 1.18e+04 & 3.81e+04\\
170 &  30.59 &  -0.11 &  115.5 &  4.3 &  7.3 &   0.8 &  19.4 & 3.54e+04 & 3.17e+04 &&   1.1 &   1.7 & 1.67e+02 & 1.90e+03\\
171 &  30.77 &  -0.01 &   94.2 &  4.6 &  5.7 &   6.8 &  41.8 & 3.17e+05 & 1.10e+06 &&   5.7 &   5.2 & 1.11e+04 & 7.59e+04\\
172 &  30.83 &  -0.18 &   51.6 &  5.9 &  3.3 &   1.8 &   8.8 & 3.06e+03 & 6.90e+03 &&   1.5 &   0.4 & 6.16e+00 & 3.12e+02\\
173 &  30.89 &  -0.60 &  102.0 &  4.4 &  7.9 &   0.7 &   8.4 & 6.67e+03 & 2.61e+03 &&   0.9 &   2.2 & 3.23e+02 & 7.74e+02\\
174 &  30.96 &   0.09 &   39.5 &  6.5 & 12.1 &   3.3 &  36.2 & 1.30e+05 & 2.22e+05 &&   2.1 &   3.3 & 1.07e+03 & 6.60e+03\\
175 &  30.97 &   0.40 &   79.7 &  5.1 &  4.7 &   2.6 &  12.5 & 2.13e+04 & 3.96e+04 &&   2.3 &   1.1 & 1.04e+02 & 1.08e+03\\
177 &  31.28 &  -0.00 &   79.7 &  5.0 &  4.9 &   3.2 &  14.7 & 2.66e+04 & 4.84e+04 &&   3.2 &   5.4 & 3.96e+03 & 1.48e+04\\
178 &  31.32 &  -0.03 &   41.0 &  6.6 & 12.1 &   3.7 &  58.6 & 2.43e+05 & 5.06e+05 &&   3.5 &   7.7 & 5.88e+03 & 4.31e+04\\
179 &  31.39 &  -0.26 &   87.9 &  4.8 &  9.1 &   1.6 &  13.6 & 1.84e+04 & 1.63e+04 &&   1.7 &   2.8 & 6.69e+02 & 4.06e+03\\
180 &  31.44 &   0.08 &  106.0 &  4.4 &  7.3 &   2.6 &  21.9 & 6.62e+04 & 8.94e+04 &&   3.0 &   3.2 & 1.77e+03 & 1.31e+04\\
181 &  31.98 &  -0.28 &   97.5 &  4.6 &  8.1 &   2.5 &  19.0 & 3.62e+04 & 3.08e+04 &&   2.6 &   7.4 & 5.03e+03 & 1.28e+04\\
182 &  32.02 &   0.06 &   96.8 &  4.6 &  8.0 &   2.1 &  37.7 & 8.93e+04 & 1.57e+05 &&   2.5 &   3.3 & 1.91e+03 & 1.54e+04\\
183 &  32.46 &   0.22 &   50.6 &  6.1 & 11.2 &   1.4 &  15.0 & 2.20e+04 & 4.02e+04 &&   1.5 &   2.0 & 2.40e+02 & 3.39e+03\\
184 &  32.70 &  -0.18 &   92.5 &  4.8 &  8.6 &   1.7 &  44.4 & 1.58e+05 & 1.38e+05 &&   1.5 &   3.0 & 7.47e+02 & 5.32e+03\\
185 &  33.38 &  -0.53 &   91.2 &  4.8 &  8.2 &   1.6 &  23.0 & 2.24e+04 & 2.81e+04 &&   1.5 &   6.3 & 2.41e+03 & 8.63e+03\\
186 &  33.36 &  -0.00 &   72.9 &  5.3 &  9.6 &   2.4 &  34.0 & 8.36e+04 & 1.06e+05 &&   2.2 &   5.5 & 3.46e+03 & 9.11e+03\\
187 &  33.44 &  -0.08 &   86.5 &  4.9 &  8.7 &   2.2 &  35.7 & 8.99e+04 & 1.14e+05 &&   2.2 &  11.4 & 9.66e+03 & 2.40e+04\\
188 &  33.79 &  -0.18 &   52.4 &  6.3 & 11.2 &   3.4 &  32.9 & 6.16e+04 & 1.16e+05 &&   3.1 &   5.6 & 2.71e+03 & 1.16e+04\\
189 &  33.66 &   0.22 &   41.7 &  6.5 & 11.6 &   1.5 &  21.3 & 2.41e+04 & 3.11e+04 &&   1.2 &   3.3 & 4.22e+02 & 3.04e+03\\
190 &  33.85 &   0.00 &   89.2 &  4.9 &  8.4 &   1.2 &  19.9 & 3.48e+04 & 2.48e+04 &&   1.2 &  12.1 & 8.33e+03 & 1.44e+04\\
191 &  33.83 &   0.07 &  105.6 &  4.7 &  7.1 &   2.2 &  30.1 & 6.69e+04 & 7.76e+04 &&   2.0 &   2.1 & 4.74e+02 & 2.83e+03\\
192 &  34.16 &  -0.10 &   88.1 &  4.9 &  8.3 &   2.1 &  21.5 & 4.01e+04 & 3.56e+04 &&   1.9 &   6.0 & 2.26e+03 & 9.18e+03\\
193 &  34.20 &   0.12 &   57.5 &  6.1 &  3.2 &   2.8 &  18.1 & 3.00e+04 & 1.07e+05 &&   2.8 &   1.4 & 1.68e+02 & 1.70e+04\\
195 &  34.36 &  -0.19 &   52.2 &  6.1 &  3.2 &   2.3 &   7.6 & 3.37e+03 & 7.86e+03 &&   2.3 &   2.7 & 5.94e+02 & 2.58e+03\\
196 &  34.76 &  -0.13 &   78.9 &  5.3 &  4.8 &   3.3 &  16.1 & 1.59e+04 & 3.37e+04 &&   3.6 &   3.7 & 1.32e+03 & 5.67e+03\\
198 &  34.99 &   0.33 &   51.8 &  6.2 &  3.1 &   1.0 &  15.7 & 2.31e+04 & 3.60e+04 &&   1.3 &   0.7 & 4.54e+01 & 8.03e+02\\
202 &  35.66 &   0.15 &   81.7 &  5.2 &  8.5 &   2.7 &  36.0 & 9.81e+04 & 1.43e+05 &&   2.2 &   2.1 & 3.41e+02 & 4.21e+03\\
203 &  35.79 &  -0.16 &   28.8 &  7.2 &  1.7 &   1.2 &  12.8 & 3.80e+03 & 7.76e+03 &&   0.9 &   0.6 & 6.14e+00 & 5.66e+02\\
204 &  35.97 &  -0.48 &   58.7 &  6.0 &  3.5 &   1.7 &   6.4 & 3.69e+03 & 4.50e+03 &&   1.6 &   2.5 & 4.44e+02 & 2.00e+03\\
205 &  36.13 &   0.66 &   77.5 &  5.4 &  4.9 &   3.6 &  31.7 & 5.58e+04 & 7.36e+04 &&   2.2 &   1.5 & 1.91e+02 & 1.30e+03\\
206 &  36.42 &  -0.10 &   54.8 &  6.3 &  3.1 &   2.0 &  17.2 & 2.74e+04 & 5.13e+04 &&   1.8 &   3.6 & 9.23e+02 & 5.00e+03\\
207 &  36.49 &  -0.11 &   78.6 &  5.4 &  4.9 &   3.1 &  25.3 & 3.51e+04 & 5.95e+04 &&   2.3 &   2.0 & 2.90e+02 & 2.10e+03\\
208 &  36.90 &  -0.07 &   79.9 &  5.3 &  5.2 &   2.2 &  13.9 & 1.81e+04 & 3.66e+04 &&   2.1 &   6.4 & 4.24e+03 & 1.58e+04\\
209 &  37.38 &   0.17 &   87.2 &  5.2 &  6.8 &   3.1 &  31.8 & 7.05e+04 & 1.24e+05 &&   2.7 &   1.4 & 1.02e+02 & 5.27e+03\\
210 &  37.49 &   0.08 &   41.9 &  6.7 &  2.4 &   1.6 &   6.4 & 1.92e+03 & 2.95e+03 &&   0.9 &   0.5 & 3.97e+00 & 2.92e+02\\
211 &  37.76 &  -0.21 &   62.8 &  6.0 &  9.8 &   3.4 &  20.8 & 5.01e+04 & 7.71e+04 &&   3.1 &   3.2 & 1.14e+03 & 8.41e+03\\
212 &  38.23 &  -0.15 &   62.9 &  5.9 &  9.3 &   2.7 &  10.8 & 2.13e+04 & 3.16e+04 &&   2.9 &   5.1 & 2.78e+03 & 1.38e+04\\
213 &  38.93 &  -0.45 &   41.7 &  6.8 &  2.5 &   1.5 &  10.8 & 9.17e+03 & 2.28e+04 &&   1.4 &   1.0 & 1.52e+02 & 1.29e+03\\
214 &  39.83 &  -0.28 &   60.3 &  6.2 &  9.6 &   3.8 &  90.8 & 7.72e+05 & 1.19e+06 &&   3.4 &  15.1 & 2.81e+04 & 1.08e+05\\
215 &  40.32 &  -0.42 &   73.5 &  5.7 &  5.0 &   0.9 &  12.5 & 8.05e+03 & 8.35e+03 &&   1.0 &   1.2 & 5.75e+01 & 5.59e+02\\
216 &  41.05 &  -0.17 &   39.8 &  7.0 &  2.2 &   1.8 &  11.4 & 8.01e+03 & 1.29e+04 &&   1.9 &   3.5 & 7.76e+02 & 3.28e+03\\
217 &  41.18 &  -0.22 &   61.1 &  6.2 &  9.0 &   1.8 &  42.6 & 1.99e+05 & 2.52e+05 &&   1.5 &   1.5 & 1.98e+02 & 3.73e+03\\
218 &  41.89 &  -0.40 &   60.3 &  6.2 &  8.9 &   1.4 &  25.3 & 5.69e+04 & 6.72e+04 &&   1.6 &   5.8 & 2.60e+03 & 9.71e+03\\
219 &  42.34 &  -0.08 &   57.5 &  6.3 &  8.9 &   3.0 &  20.6 & 5.06e+04 & 6.33e+04 &&   2.7 &   3.5 & 1.11e+03 & 4.59e+03\\
220 &  42.15 &  -0.60 &   67.2 &  6.0 &  4.5 &   2.2 &  15.9 & 2.84e+04 & 5.52e+04 &&   1.7 &   0.3 & 4.58e+00 & 3.59e+02\\
221 &  42.72 &  -0.35 &   62.6 &  6.3 &  3.8 &   3.8 &  16.9 & 1.97e+04 & 5.74e+04 &&   3.5 &   1.5 & 1.46e+02 & 1.11e+03\\
223 &  43.17 &  -0.52 &   57.9 &  6.4 &  3.6 &   1.2 &   7.1 & 2.56e+03 & 6.79e+03 &&   1.5 &   0.5 & 2.22e+01 & 1.34e+03\\
224 &  44.38 &  -0.22 &   61.2 &  6.2 &  7.7 &   4.0 &  46.1 & 1.91e+05 & 4.59e+05 &&   2.3 &   0.8 & 4.67e+01 & 2.68e+03\\
225 &  45.44 &   0.07 &   59.9 &  6.4 &  8.1 &   4.4 &  47.0 & 1.66e+05 & 4.43e+05 &&   3.0 &   3.6 & 2.03e+03 & 2.03e+04\\
226 &  46.32 &  -0.20 &   55.0 &  6.4 &  7.7 &   3.4 &  13.6 & 1.42e+04 & 5.82e+04 &&   2.5 &   1.5 & 1.41e+02 & 3.43e+03\\
227 &  47.05 &   0.26 &   57.6 &  6.6 &  7.9 &   2.1 &  19.8 & 2.15e+04 & 6.81e+04 &&   1.7 &   2.2 & 2.63e+02 & 1.29e+04\\
229 &  47.55 &  -0.54 &   59.1 &  6.4 &  7.2 &   1.7 &  17.7 & 1.97e+04 & 1.95e+04 &&   1.6 &   3.7 & 5.69e+02 & 3.32e+03\\
232 &  48.83 &   0.14 &   52.6 &  6.7 &  7.7 &   1.9 &  32.0 & 3.42e+04 & 1.28e+05 &&   1.5 &   2.0 & 2.25e+02 & 2.98e+03\\
234 &  49.74 &  -0.52 &   68.1 &  6.5 &  5.5 &   1.1 &  11.9 & 9.88e+03 & 1.89e+04 &&   0.9 &   2.7 & 4.92e+02 & 2.61e+03\\
236 &  50.83 &   0.25 &   42.3 &  7.0 &  2.9 &   2.0 &   7.3 & 2.62e+03 & 7.69e+03 &&   1.6 &   0.7 & 1.79e+01 & 7.96e+02\\
237 &  51.33 &  -0.04 &   54.7 &  6.7 &  6.2 &   2.5 &  31.3 & 6.05e+04 & 9.99e+04 &&   1.9 &   0.8 & 4.53e+01 & 4.28e+03\\
238 &  52.30 &  -0.06 &   51.1 &  6.8 &  6.1 &   1.6 &  12.3 & 1.38e+04 & 1.90e+04 &&   1.6 &   5.2 & 1.63e+03 & 6.23e+03\\
240 &  53.17 &  -0.25 &   62.7 &  6.8 &  5.1 &   2.3 &   9.0 & 8.71e+03 & 1.28e+04 &&   1.9 &   1.3 & 1.32e+02 & 1.37e+03\\
241 &  53.43 &   0.07 &   23.3 &  7.7 &  1.5 &   1.1 &  15.9 & 5.19e+03 & 2.26e+04 &&   0.9 &   0.2 & 6.26e+00 & 2.72e+02\\
242 &  54.12 &  -0.07 &   39.1 &  7.2 &  7.0 &   2.5 &  12.8 & 1.24e+04 & 5.03e+04 &&   2.3 &   3.5 & 1.84e+03 & 1.95e+04\\
243 &  54.66 &   0.81 &   32.8 &  7.5 &  7.7 &   2.9 &  28.2 & 4.33e+04 & 1.12e+05 &&   2.8 &   7.3 & 5.11e+03 & 2.17e+04\\
\enddata
\end{deluxetable} 
\end{landscape}

\end{document}